\documentclass[twocolumn,english,aps,prb,showpacs,showkeys]{revtex4}
\usepackage[T1]{fontenc}
\usepackage[latin9]{inputenc}
\usepackage{color}
\usepackage{amsmath}
\usepackage{amssymb}
\usepackage{graphicx}
\usepackage{esint}

\makeatletter
\@ifundefined{textcolor}{}
{%
 \definecolor{BLACK}{gray}{0}
 \definecolor{WHITE}{gray}{1}
 \definecolor{RED}{rgb}{1,0,0}
 \definecolor{GREEN}{rgb}{0,1,0}
 \definecolor{BLUE}{rgb}{0,0,1}
 \definecolor{CYAN}{cmyk}{1,0,0,0}
 \definecolor{MAGENTA}{cmyk}{0,1,0,0}
 \definecolor{YELLOW}{cmyk}{0,0,1,0}
 }

\usepackage{babel}

\newcommand{\1}{{\bf \scriptstyle 1}\!\!{1}}

\usepackage{babel}

\usepackage{babel}

\usepackage{babel}

\makeatother

\usepackage{babel}
\begin{document}

\title{Giant Faraday effect due to Pauli exclusion principle in 3D topological
insulators}

\author{Hari P. Paudel}

\author{Michael N. Leuenberger}

\email{michael.leuenberger@ucf.edu}

\affiliation{NanoScience Technology Center and Department of Physics, 12424 Research
Parkway Suite 400, Orlando, Florida 32826, United States}
\begin{abstract}
Experiments using ARPES, which is based on the photoelectric effect,
show that the surface states in 3D topological insulators (TI) are
helical. Here we consider Weyl interface fermions due to band inversion
in narrow-bandgap semiconductors, such as Pb$_{1-x}$Sn$_{x}$Te.
The positive and negative energy solutions can be identified by means
of opposite helicity in terms of the spin helicity operator in 3D
TI as $\hat{h}_{\textrm{TI}}=\left(1/\left|p_{\bot}\right|\right)\beta\left(\boldsymbol{\sigma}_{\perp}\times\boldsymbol{p}_{\perp}\right)\cdot\boldsymbol{\hat{z}}$,
where $\beta$ is a Dirac matrix and $\boldsymbol{\hat{z}}$ points
perpendicular to the interface. Using the 3D Dirac equation and bandstructure
calculations we show that the transitions between positive and negative
energy solutions, giving rise to electron-hole pairs, obey strict
optical selection rules. In order to demonstrate the consequences
of these selection rules, we consider the Faraday effect due to Pauli
exclusion principle in a pump-probe setup using a 3D TI double interface
of a PbTe/Pb$_{0.31}$Sn$_{0.69}$Te/PbTe heterostructure. For that
we calculate the optical conductivity tensor of this heterostructure,
which we use to solve Maxwell's equations. The Faraday rotation angle
exhibits oscillations as a function of probe wavelength and thickness
of the heterostructure. The maxima in the Faraday rotation angle are
of the order of millirads. 
\end{abstract}

\pacs{78.67.-n,78.67.Hc,78.67.Wj,71.15.Mb}

\keywords{topological insulator, electron-hole pair, selection rule, density
functional theory.}

\maketitle

\section{Introduction}

The 3D TI is a new state of matter on the surface or at the interface
of narrow-bandgap materials where topologically protected gapless
surface/interface states appear within the bulk insulating gap.\cite{key-1,key-2,key-11,Hsieh:2009,Hajlaoui:2012,key-6-1,key-7-1,key-5-1}
These states are characterized by the linear excitation energy of
massless Weyl fermions. The spins of the Kramers partners are locked
at a right angle to their momenta due to the Rashba spin-orbit coupling,\cite{key-21-1}
protecting them against perturbation and scattering.\cite{key-1,key-2,key-3,key-4}
Because of the presence of a single Dirac cone with fixed spin direction
at the \emph{surface}, the main feature of strong TIs,\cite{key-8,key-9}
the materials Bi$_{2}$Se$_{3}$ and Bi$_{2}$Te$_{3}$ are currently
being widely studied.\cite{key-10,key-11,key-5-1}

The heterostructures of compound semiconductors such as Bi$_{1-x}$Sb$_{x}$
and Pb$_{1-x}$Sn$_{x}$Te exhibit a strong topological phase.\cite{key-4}
In Bi$_{1-x}$Sb$_{x}$, the $L^{+}$ and $L{}^{-}$ bands cross at
$x=0.04$. The pure PbTe has inverted bands at the band gap extrema
with respect to SnTe. In Pb$_{1-x}$Sn$_{x}$Te, initially increasing
the concentration of Sn leads to a decreasing band gap. At around
$x=0.35$, the bands cross and the gap reopens for $x>0.35$ with
even parity $L^{+}$ band and odd parity $L{}^{-}$ band being inverted
with respect to each other.\cite{key-12} The band inversion between
PbTe and SnTe results in \emph{interface} states,\cite{key-13,key-14,key-15}
which can be described by the Weyl equation.\cite{key-16} 
\begin{figure}
\textbf{\includegraphics[width=8.5cm]{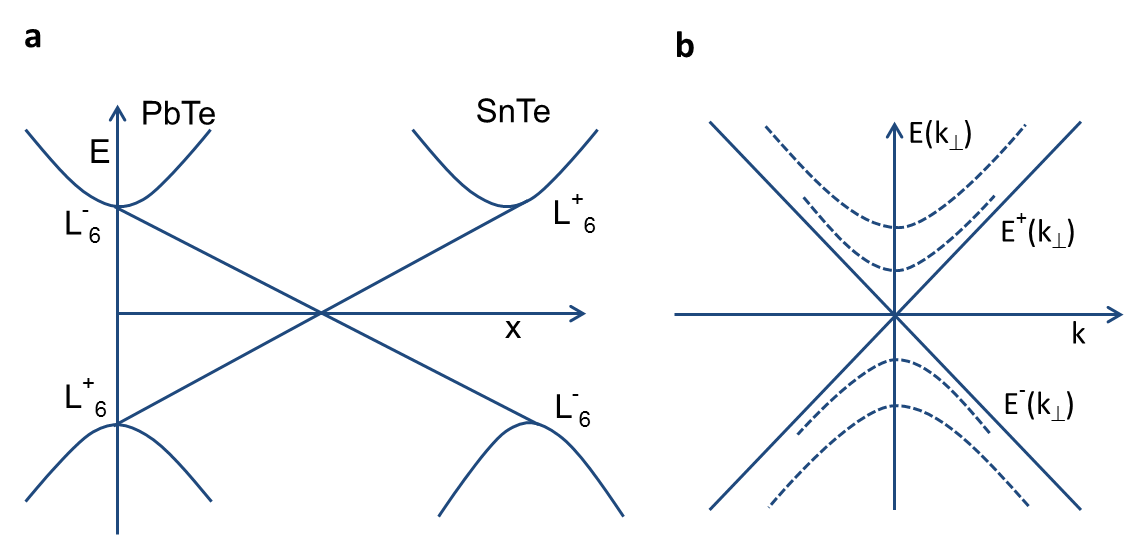}}\caption{(a) Band inversion in between two end members in Pb$_{1-x}$Sn$_{x}$Te.
(b) Energy spectrum of the inverted contact. The solid lines are Weyl
states and dashed lines are additional branches that appear for contact
thickness $l>l_{o}$. \label{fig:Band-inversion-in}}
\end{figure}

Here we investigate the giant Faraday effect due to Pauli exclusion
principle and the strict optical selection rules governing the low
energy excitation of electron-hole pairs around a Dirac point in a
3D TI. Due to interference effects, the Faraday rotation angle exhibits
oscillations as a function of probe wavelength and thickness of the
slab material on either side of the 3D TI double interface of a PbTe/Pb$_{0.31}$Sn$_{0.69}$Te/PbTe
heterostructure. The maxima in the Faraday rotation angle are in the
mrad regime. We find that in 3D TIs both interband transitions (between
positive and negative energy solutions) and intraband transitions
(within the same energy solutions) are allowed. Note that the selection
rules obtained here are different from the selection rules in ARPES
experiments, which record the number of photoelectrons as a function
of kinetic energy and emission angle with respect to the sample surface.
A number of experiments have shown the existence of the helical surface
states in 3D TI.\cite{key-10,key-11} As an example, we consider the
alloy Pb$_{1-x}$Sn$_{x}$Te, which has topologically nontrivial interface
states under appropriate doping level. Our results are qualitatively
valid for all strong 3D TIs. Pb$_{1-x}$Sn$_{x}$Te has a rocksalt
type crystal lattice with four non-equivalent L points located in
the center of the hexagonal facets on {[}111{]} axis. The valence
and conduction band edges are derived from the hybridized p-type and
s-type orbitals at the $L$ point.\cite{key-17} Its end species have
inverted band character, $L{}^{+}$ character of PbTe band switches
to $L{}^{-}$ character of SnTe band and vice versa as shown in Fig.
\ref{fig:Band-inversion-in}. The Brillouin zone of the Pb$_{1-x}$Sn$_{x}$Te
crystal has eight hexagonal faces each with center at the $L$ point.
Two faces lying diametrically opposite are equivalent. As a result,
the band inversion happens at four distinct Dirac points. The crystal
possesses a mirror symmetry. Therefore, it is a distinct class of
3D TI where surface states are protected by mirror symmetry.\cite{Hasan}
We choose the $z$-axis to point in direction of the gradient of the
concentration $\nabla x$. At the two band extrema, the low energy
Hamiltonian is described by a 3D relativistic Dirac equation whose
solutions are localized near the $z=0$ plane where the band crossing
occurs, which defines the interface. Dispersion is nearly linear owing
to the large band velocities of $v_{\perp}=8\times10^{5}$ m/s and
$v_{\parallel}=2.24\times10^{5}$ m/s with a small gap.\cite{key-16}
Such properties result in a small localization length $l{}_{o}$ of
the interface wave functions along the $z$-axis. Due to the absence
of a center of inversion, a Rashba-type spin-orbit coupling is present,
which is automatically taken into account through the Dirac equation.
We also present the details of our ab-initio calculation of the bandstructure
in the supercell Brillouin zone obtained by doubling the lattice parameters
in each direction. Analysis of the alloy band structures is usually
complicated due to folding of the bands from neighboring Brillouin
zones, making it difficult to map the calculated bandstructures onto
the bandstructures obtained from momentum-resolving experiments. The
analysis is further complicated by the presence of impurity bands
inside the normal bulk energy gap. The interface states sometimes
overlap with bulk energy states. Therefore, we unfold the band structures
along the {[}111{]} direction in order to shift the band crossing
from the $\Gamma$ point, as seen in the supercell Brillouin zone,
to the $L$ point in the primitive cell Brillouin zone.\cite{key-18,key-19}

\begin{figure}
\includegraphics[width=5cm,height=5cm]{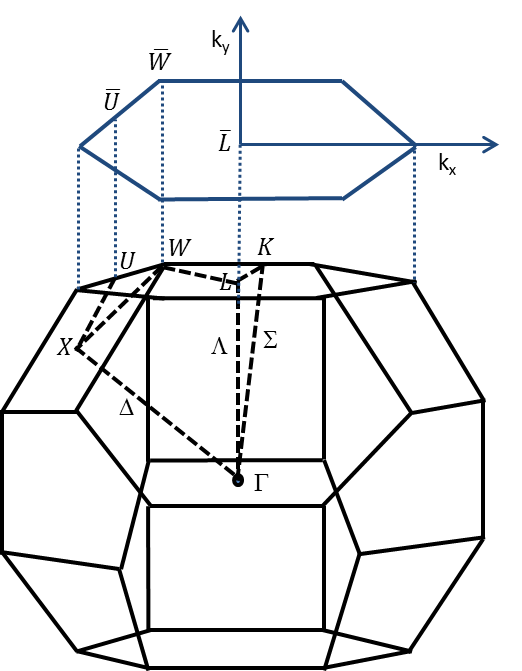}\caption{Brillouin zone for rocksalt type crystal with space group Fm$\bar{3}$m.
There are four inequivalent $L$ points at the center of the faces
on the surface of the Brillouin zone. The growth direction is along
{[}111{]} and is chosen to point along the $z$-axis. With the appropriate
level of doping by Sn atoms in PbTe, band gap goes to zero at L point
with a linear excitation energy that traces out a cone in the 2D Fermi
surface parallel to the face of Brillouin zone that is perpendicular
to the growth direction.}
\end{figure}

We developed a method of the Faraday rotation of a single photon due to Pauli exclusion
principle for a topologically trivial quantum dot\cite{Leuenberger:2005, Leuenberger:2006}
and for a 3D TI quantum dot.\cite{Paudel:2013} The
proposed method can be used for entangling remote excitons, electron
spins, and hole spins. We showed that this entanglement can be used
for the implementation of optically mediated quantum teleportation
and quantum computing. 

\begin{figure}
\includegraphics[width=7.5cm]{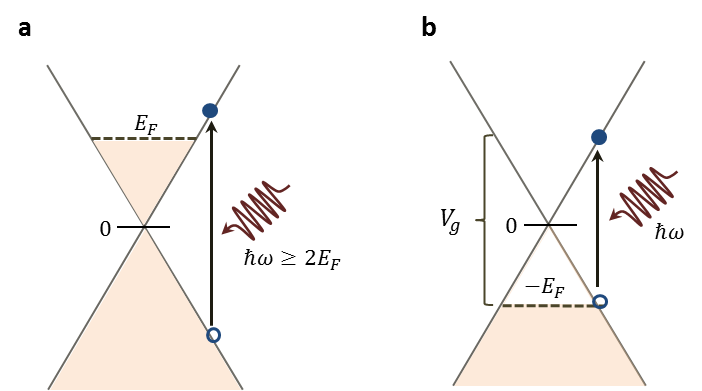}\caption{Transparency region for the optical excitation of the Weyl fermion.
The light yellow color (online) represents the filled Fermi sea of
the electrons. The zero energy is defined by the apex of the Dirac
cone. \textbf{a} With the photon energy of $\hbar\omega\geq2E_{F}$
a Weyl fermion can be excited. There are no transitions for a photon
energy below $2E_{F}$. The Fermi level is measured from the zero
of the energy. \textbf{b} The Fermi level can be shifted below the
Dirac node by applying a gate voltage $V_{g}\geq E_{F}$. Then a Weyl
fermion can be excited with a photon energy of $\hbar\omega\geq\left|E_{F}\right|$.\label{OpticalExcitation}}
\end{figure}

\begin{figure}
\includegraphics[width=7.5cm]{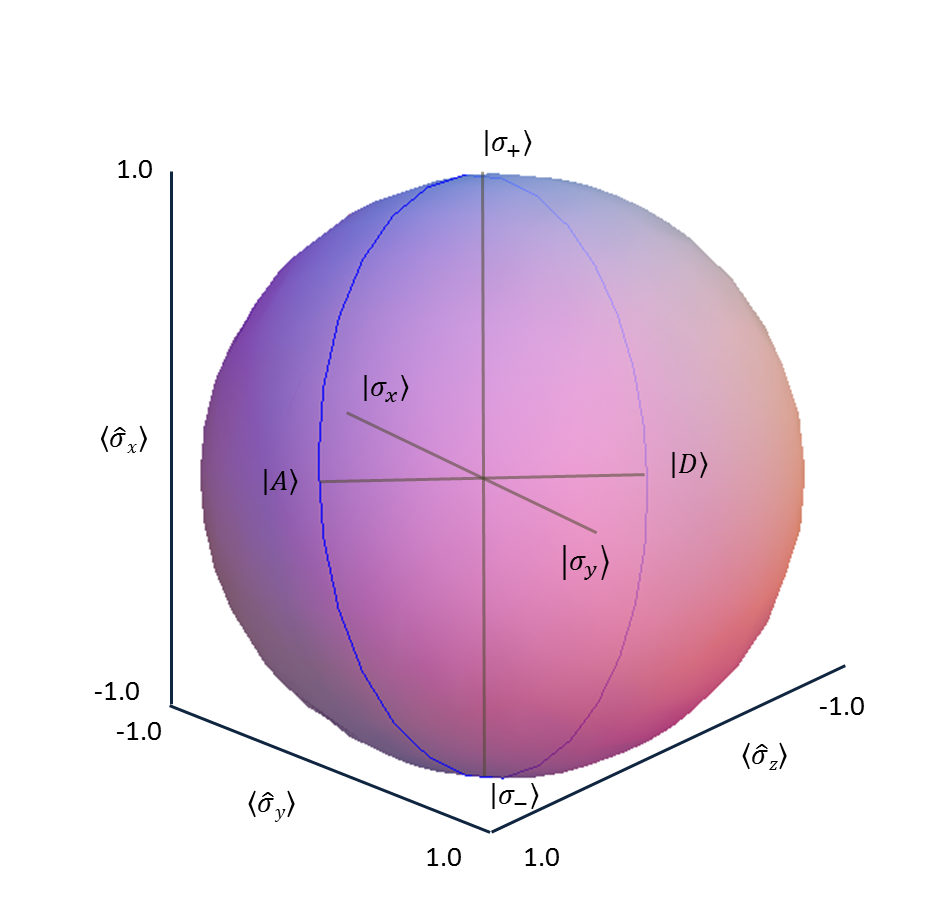}\caption{Illustration of the diagonal, $\left|D\right\rangle $ and anti-diagonal,
$\left|A\right\rangle $ polarization in a Poincare sphere. $\left\langle \hat{\sigma}_{x}\right\rangle $,
$\left\langle \hat{\sigma}_{y}\right\rangle $ and $\left\langle \hat{\sigma}_{z}\right\rangle $
are the expectation values of the Pauli matrices $\sigma_{x}=\left(\protect\begin{array}{cc}
0 & 1\protect\\
1 & 0
\protect\end{array}\right)$, $\sigma_{y}=\left(\protect\begin{array}{cc}
0 & -i\protect\\
i & 0
\protect\end{array}\right)$ and $\sigma_{z}=\left(\protect\begin{array}{cc}
1 & 0\protect\\
0 & -1
\protect\end{array}\right)$, respectively. $\left|\sigma_{x}\right\rangle $ and $\left|\sigma_{y}\right\rangle $
represent the $x$- and $y$-polarization states and $\left|\sigma_{+}\right\rangle $
and $\left|\sigma_{-}\right\rangle $ represent the left and right
circular polarization states of the photon\label{PoincareSphere}.
\label{fig:Poincare_sphere}}
\end{figure}

Here we investigate the Faraday effect due to the Pauli exclusion
principle for a 3D TI double interface of a PbTe/Pb$_{0.31}$Sn$_{0.69}$Te/PbTe
heterostructure. This Faraday effect is completely different from
the Faraday effect due to an external magnetic field, which was presented
in Ref. \onlinecite{TseMacDonald} for a thin film of a 3DTI where
a gap was opened by breaking the time reversal symmetry through a
magnetic field. The Faraday effect presented here arises from the
polarization of electron-hole (e-h) pairs that are excited by means
of a linearly polarized laser pump beam. A laser probe beam with energy
below twice the absolute value of the Fermi energy measured from the
Dirac point cannot be absorbed due to the absence of charge carriers
in this energy regime. The excitation of the Weyl fermion can happen
when a photon has an energy of $\hbar\omega\geq2E_{F}$ as shown in
the Fig. \ref{OpticalExcitation}\textbf{a}. There are no interband
transitions with the photon energy less than $2E_{F}$. A gate voltage
can also be applied to shift the Fermi level below the Dirac node\textbf{\cite{Kim}}.
Fig. \ref{OpticalExcitation}\textbf{b }shows the scheme of the gate-induced
shift in the Fermi level. In the Figure photon of energy $\hbar\omega\geq\left|E_{F}\right|$
can excite a Weyl fermion. We call this energy regime the transparency
region, in order to avoid confusion with a bandgap in a gapped semiconductor
material. When $x$- and $y$-linearly polarized e-h pairs are present,
a probe beam linearly polarized along the diagonal direction $x+y$
experiences a Faraday rotation on the Poincare sphere as shown in
Fig. \ref{fig:Poincare_sphere}. The resulting Faraday rotation angle
is giant and of the order of mrad. It exhibits oscillations as a function
of the slab thickness of the two PbTe layers of the PbTe/Pb$_{0.31}$Sn$_{0.69}$Te/PbTe
heterostructure containing two interfaces (see Fig. \ref{fig:layers}).
The Pb$_{0.31}$Sn$_{0.69}$Te is 10 nm thick in order to introduce
a gap for three out of the four L-points, as described below. The
Faraday effect results then only from the excitation of e-h pairs
at a single L-point.

The paper is organized as follows. In Sec.~\ref{sec:model} we present
the analytical derivation of the Weyl solution of the Dirac equation
that describes the level crossing at the $L$ point. Using the Rashba
spin-orbit Hamiltonian, we derive the helicity operator for 3D topological
insulators in Sec.~\ref{sec:helicity}. Sec.~\ref{sec:matrix_elements}
is devoted to the evaluation of the optical transition matrix elements.
The resulting optical selection rules are discussed in Sec.~\ref{sec:selection_rules}.
In order to obtain a quantitative result for the optical transition
matrix elements, we perform a bandstructure calculation of the alloy
Pb$_{1-x}$Sn$_{x}$Te in Sec.~\ref{sec:bandstructure}. The Sec.\ref{Faraday-Effect for 3D TIs}
is devoted to the explicit derivation of the Faraday rotation effect
and calculation of the Faraday rotation angle in the PbTe/Pb$_{0.31}$Sn$_{0.69}$Te/PbTe
heterostructure. 

\section{Model based on Dirac equation}

\label{sec:model} The energy spectrum of Pb$_{1-x}$ Sn$_{x}$Te
near the $L{}_{6}^{\mp}$ band crossing is described within the $\mathbf{k}\cdot\mathbf{p}$
perturbation theory by the two-band Dirac Hamiltonian\cite{Nimtz:1983}
\begin{equation}
H=\left(\begin{array}{cc}
\Delta(z) & v_{\Vert}\sigma_{z}\hat{p}_{z}+v_{\bot}\boldsymbol{\sigma}_{\bot}\cdot\mathbf{\hat{p}}_{\bot}\\
v_{\Vert}\sigma_{z}\hat{p}_{z}+v_{\bot}\boldsymbol{\sigma}_{\bot}\cdot\mathbf{\hat{p}}_{\bot} & -\Delta(z)
\end{array}\right),\label{eq:1}
\end{equation}
where $\mathbf{\boldsymbol{\sigma}}$ are the Pauli matrices, $\mathbf{\hat{p}}=-i\hbar\boldsymbol{\nabla}$
is the momentum operator and $\Delta(z)=\varepsilon_{g}\left(z\right)/2$
is the gap energy parameter with symmetry $\Delta\left(z\right)=-\triangle\left(-z\right)$.
$\boldsymbol{\sigma}_{\bot}=(\sigma_{x},\sigma_{y})$ and $\mathbf{\hat{p}}_{\bot}=(\hat{p}_{x},\hat{p}_{y})$
denote the Pauli matrices and momenta in the interface plane, respectively.
The transverse and longitudinal velocities are determined by $v_{\bot}=P_{\bot}/m_{0}$
and $v_{\Vert}=P_{\Vert}/m_{0}$, where $P_{\bot}$ and $P_{\Vert}$
are the transverse and longitudinal Kane interband matrix elements,
respectively. $m_{0}=9.10938188\times10^{-31}$ kg is the free electron
mass. The inhomogeneous structure is synthesized by changing the composition
along one of the {[}111{]} axes, whose symmetry breaking leads to
a single Dirac cone in the chosen direction,\cite{key-23-1} thereby
recovering the Z$_{2}$ strong topological insulator phase. The direction
of the gradient of the concentration $\boldsymbol{\nabla}x$ defines
our $z$-axis. After the unitary transformation of the Hamiltonian
$H=UH'U^{\dagger}$ using 
\begin{equation}
U=\frac{1}{\sqrt{2}}\left(\begin{array}{cc}
\1 & i\1\\
i\1 & \1
\end{array}\right),
\end{equation}
the time-independent Dirac equation $H'\Phi_{\pm}^{'}=\left(\varepsilon-\varphi\left(z\right)\right)\Phi_{\pm}^{'}$
can be written as 
\begin{align}
 & \left(\begin{array}{cc}
0 & i\Delta+v_{\Vert}\sigma_{z}\hat{p}_{z}+v_{\bot}\boldsymbol{\sigma}_{\bot}\cdot\mathbf{\hat{p}}_{\bot}\\
-i\Delta+v_{\Vert}\sigma_{z}\hat{p}_{z}+v_{\bot}\boldsymbol{\sigma}_{\bot}\cdot\mathbf{\hat{p}}_{\bot} & 0
\end{array}\right)\nonumber \\
 & \times\left(\begin{array}{c}
\phi{'}^{L^{-}}\\
\phi{'}^{L^{+}}
\end{array}\right)=\left(\varepsilon-\varphi\left(z\right)\right)\left(\begin{array}{c}
\phi{'}^{L^{-}}\\
\phi{'}^{L^{+}}
\end{array}\right)\label{eq:3}
\end{align}
where $\phi{'}^{L^{-}}$ and $\phi{'}^{L^{+}}$ are the two-component
spinors of the $L^{-}$ and the $L^{+}$ band, respectively. The potential
$\varphi\left(z\right)$ (work function) describes the variation of
the gap center. For simplicity we consider the case $\varphi(z)=0$.
From Eq. (\ref{eq:3}), the two-component spinor $\phi{'}^{L^{\pm}}$
satisfies 
\begin{equation}
\left(p^{2}+U_{\pm}\left(z,\sigma_{z}\right)-\varepsilon^{2}\right)\phi{'}^{L^{\pm}}=0
\end{equation}
where $U_{\pm}\left(z,\sigma_{z}\right)=\Delta^{2}\pm\hbar v_{\parallel}\sigma_{z}\frac{\partial\Delta}{\partial z}$.
In its origin, the linear Weyl spectrum $\varepsilon_{o}^{\pm}\left(k_{\perp}\right)=\pm\hbar v_{\perp}k_{\perp}$at
$k_{\perp}=0$ is approximately equal to the soliton spectrum in the
1D Peierl's insulator. This implies that $\Delta\left(z\right)$ can
be chosen to be $\Delta\left(z\right)=\Delta\left(\infty\right)\tanh\left(z/l\right)$.
Interface states are localized along the $z$-axis with the localization
length $l_{o}=\hbar v_{\Vert}/\Delta\left(\infty\right)$. For $l_{o}<l$,
additional branches with finite mass appear. There are several solutions
at $\varepsilon^{2}>\Delta^{2}\left(\infty\right)$ which are localized
at the contact. For $l_{o}>l$, only Weyl solutions exist. We focus
on the case when $l_{o}>l$. Then we have only zero-energy solutions,
which correspond to the Weyl states and are given by \cite{key-16}
\begin{equation}
\Phi_{\pm}^{'}=C\left(\begin{array}{c}
\pm e^{-\frac{i\theta}{2}}\\
0\\
0\\
e^{\frac{i\theta}{2}}
\end{array}\right)e^{-\frac{1}{\hbar v_{\parallel}}\overset{z}{\underset{0}{\int}}\Delta\left(z^{'}\right)dz^{'}+i\mathbf{k}_{\perp}\cdot\mathbf{r}}\label{eq:5}
\end{equation}
where C is a normalization constant, $\mathbf{k}_{\perp}=\left(k_{x},\: k_{y},\:0\right)$
and $e^{\mp i\theta}=\frac{k_{x}\mp ik_{y}}{k_{\perp}}$. These solutions
have eigenenergies $\varepsilon_{o}^{\pm}\left(k_{\perp}\right)=\pm\hbar v_{\perp}k_{\perp}$.
For $\Delta\left(z\right)$ to vanish at the inverted contact, it
can be seen from Eq. (\ref{eq:3}) that $\phi{'}_{\pm}^{L^{-}}$ and
$\phi{'}_{\pm}^{L^{+}}$ must have only non-zero spin down and spin
up components, respectively. Each spinor at $L^{\mp}$ band can be
represented with the spin up states from the $L^{-}$ band and spin
down states from the $L^{+}$ band for both the positive and the negative
energies. The motion of the particle at the inverted contact is separated
into free motion in the $xy$-plane and confinement along the $z$-axis.
A remarkable property of Eq. (\ref{eq:3}) is the presence of the
zero mode (Weyl mode) localized around $z=0$. It is this mode that
has a locked spin structure. In order to understand the direction
in which the 4-spinors point, we have to transform the solutions back
to the original basis of the Hamiltonian in Eq. (\ref{eq:1}). After
the back transformation $\Phi_{\pm}=U\Phi_{\pm}^{'}$, the Weyl solutions
are 
\begin{equation}
\Phi_{\pm}=Ce^{\pm i\frac{\pi}{4}}\left(\begin{array}{c}
\pm e^{-i\frac{\left(\theta\pm\pi/2\right)}{2}}\\
\pm e^{i\frac{\left(\theta\pm\pi/2\right)}{2}}\\
e^{-i\frac{\left(\theta\mp\pi/2\right)}{2}}\\
e^{i\frac{\left(\theta\mp\pi/2\right)}{2}}
\end{array}\right)e^{-\frac{1}{\hbar v_{\parallel}}\overset{z}{\underset{0}{\int}}\Delta(z^{'})dz^{'}+i\mathbf{k}_{\perp}\cdot\mathbf{r}}\label{eq:6}
\end{equation}
where C is a normalization constant. These solutions have eigenenergies
$\varepsilon_{o}^{\pm}\left(k_{\perp}\right)=\pm\hbar v_{\perp}k_{\perp}$
and are \textit{helical}. At this time, it is useful to introduce
the notation 
\begin{equation}
\Phi_{\pm}=\left(\begin{array}{c}
\phi_{\pm}^{L^{-}}\\
\phi_{\pm}^{L^{+}}
\end{array}\right)=\left(\begin{array}{c}
\chi_{\pm}^{L^{-}}\\
\chi_{\pm}^{L^{+}}
\end{array}\right)F(\mathbf{r})=\chi_{\pm}F(\mathbf{r}),\label{eq:7}
\end{equation}
where $\chi_{\pm}$ is the four-spinor consisting of the two-spinors
$\chi_{\pm}^{L^{-}}$ and $\chi_{\pm}^{L^{+}}$ are two-spinors, and
$F(\mathbf{r})=Ce^{-\frac{1}{\hbar v_{\parallel}}\overset{z}{\underset{0}{\int}}\Delta(z^{'})dz^{'}+i\mathbf{k}_{\perp}\cdot\mathbf{r}}$.
We define $F(z)=e^{-\frac{1}{\hbar v_{\parallel}}\overset{z}{\underset{0}{\int}}\Delta(z^{'})dz^{'}}$.

\section{Helicity operator}

\label{sec:helicity}

\begin{figure}
\includegraphics[width=8.5cm]{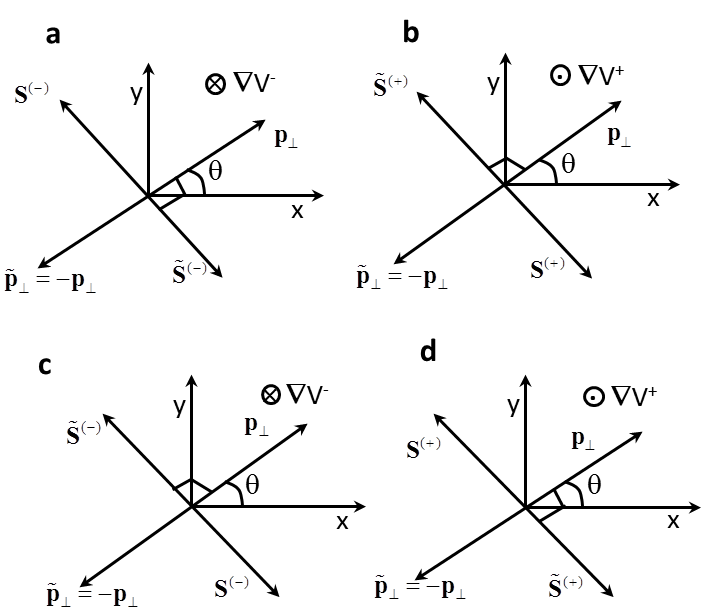}\caption{Effect of Rashba spin-orbit coupling. Spin vector $\boldsymbol{S}^{\left(-\right)}$
($\boldsymbol{\tilde{S}}^{\left(-\right)}$) in the $L^{-}$ band
and spin vector $\boldsymbol{S}^{\left(+\right)}$ ($\boldsymbol{\tilde{S}}^{\left(+\right)}$)
in the $L^{+}$ band are perpendicular to both the $z$-axis and $\boldsymbol{p}_{\perp}$
($\boldsymbol{\tilde{p}}_{\perp}$) for Weyl interface states (Weyl
Kramers partner states). (a) and (b) ((c) and (d)) correspond to positive
(negative) eigenenergy. \label{SpinLocking}}
\end{figure}

We show in this section that it is possible to clearly identify the
positive and negative energy solutions by means of a spin helicity
operator. In the representation shown in Eq. (\ref{eq:7}), the spin
directions reveal themselves clearly: the spins of the two-spinors
$\chi_{\pm}^{L^{-}}$ and $\chi_{\pm}^{L^{+}}$ point perpendicular
to $\mathbf{k}_{\perp}$ owing to the $\mp\pi/2$ shifts. For an asymmetric
scalar potential $V$ applied to a semiconductor heterostructure,
the inversion symmetry is broken, which leads to the Rashba spin-orbit
coupling.\cite{key-21-1,key-20} Here in the case of the interface
of a 3D TI we have antisymmetric potentials $V^{\mp}=\pm\Delta$,
which correspond to the diagonal elements of the Hamiltonian $H$
and whose signs depend on the band $L^{\mp}$. This results in a band-dependent
Rashba spin-orbit coupling. For the positive (negative) solutions
the Rashba spin-orbit coupling has the form $H_{R}=\mp\lambda_{R}\boldsymbol{\sigma}\cdot\left(\boldsymbol{p}\times\boldsymbol{\nabla}V^{\mp}\right)=\mp\lambda_{R}\boldsymbol{\nabla}V^{\mp}\cdot\left(\boldsymbol{\sigma}\times\boldsymbol{p}\right)$,
($\left(-\right)$ sign for positive energy solutions and $\left(+\right)$
sign for negative energy solutions), where $\lambda_{R}\geq0$ is
the Rashba spin-orbit coupling constant. It is to be noted that in
both cases each spin $S^{\left(\mp\right)}$ is perpendicular both
to the momentum and to the potential gradient direction, i.e. the
$z$-axis (see Fig. \ref{SpinLocking}). Our findings are consistent
with the spin density functional calculations (DFT).\cite{key-22}

In order to determine the Kramers partners explicitly, we rotate the
phase of each of the two-spinor wavefunction by an angle $\pi$ in
the 2D interface plane, yielding $\phi_{\pm}^{L^{-}}\left(\theta+\pi\right)=e^{\mp i\frac{\pi}{4}}\left(\begin{array}{c}
-e^{-i\frac{\left(\theta\mp\pi/2\right)}{2}}\\
-e^{i\frac{\left(\theta\mp\pi/2\right)}{2}}
\end{array}\right)F^{*}(\mathbf{r})$ and $\phi_{\pm}^{L^{+}}\left(\theta+\pi\right)=e^{\mp i\frac{\pi}{4}}\left(\begin{array}{c}
\pm e^{-i\frac{\left(\theta\pm\pi/2\right)}{2}}\\
\pm e^{i\frac{\left(\theta\pm\pi/2\right)}{2}}
\end{array}\right)F^{*}(\mathbf{r})$. Their spin and momentum direction are flipped by an angle $\pi$
(Fig. \ref{SpinLocking}). This provides a theoretical hallmark of
Kramers partners in 3D TI.

Helical properties of solutions given by Eq. (\ref{eq:6}) apply to
all 3D TIs. In the case of free neutrinos in 3D space, the standard
helicity operator $\hat{h}_{n}=-\left(1/\left|p_{\bot}\right|\right)\boldsymbol{\sigma}\cdot\boldsymbol{p}$
for the spin $\mathbf{S}=\hbar\boldsymbol{\sigma}/2$ can be used.
Similarly, in the case of graphene the helicity for the pseudospin
is given by $\hat{h}_{g}=-\left(1/\left|p_{\bot}\right|\right)\boldsymbol{\sigma}\cdot\boldsymbol{p}$.
However, in the case of 3D TI this definition is not useful, because
the spin points perpendicular to the momentum. Therefore, since we
know that the Rashba spin-orbit coupling is responsible for the helicity
in 3D TIs, we define the 3D TI helicity operator as 
\begin{eqnarray}
\hat{h}_{\textrm{TI}} & = & \left(1/\left|p_{\bot}\right|\right)\left(\begin{array}{cc}
\left(\boldsymbol{\sigma}_{\perp}\times\boldsymbol{p}_{\perp}\right)\cdot\boldsymbol{\hat{z}} & 0\\
0 & -\left(\boldsymbol{\sigma}_{\perp}\times\boldsymbol{p}_{\perp}\right)\cdot\boldsymbol{\hat{z}}
\end{array}\right)\nonumber \\
 & = & \left(1/\left|p_{\bot}\right|\right)\beta\left(\boldsymbol{\sigma}_{\perp}\times\boldsymbol{p}_{\perp}\right)\cdot\boldsymbol{\hat{z}}\label{eq:8}
\end{eqnarray}
where $\boldsymbol{\sigma}_{\perp}=\left\{ \sigma_{x},\sigma_{y}\right\} $
is the 2D vector of Pauli matrices in the $xy$-plane and $\beta=\left(\begin{array}{cc}
\1 & 0\\
0 & -\1
\end{array}\right)$ is a Dirac matrix. Note that the $+$ and $-$ signs in front of
the diagonal terms are due to the direction of $\boldsymbol{\nabla}V^{\mp}$
and thus a direct consequence of the Rashba spin-orbit coupling. The
eigenfunctions of the operator $\hat{h}_{\mathrm{TI}}$ are the 4-spinor
wavefunctions given by the Eq. (\ref{eq:6}) with the eigenvalues
$\left(+1\right)$ for the positive energy solution and $\left(-1\right)$
for the negative energy solution, i.e. $\hat{h}_{\mathrm{TI}}\Phi_{\pm}=\left(\pm1/2\right)\Phi_{\pm}$.
$\hat{h}_{\mathrm{TI}}$ commutes with the Hamiltonian in Eq.~(\ref{eq:1}).
This provides the possibility to write an effective 2D Hamiltonian
for the Weyl fermions on the surface of 3D topological insulators,
i.e. 
\begin{equation}
H_{2D}=\hbar v\left(\begin{array}{cc}
\left(\boldsymbol{\sigma}_{\perp}\times\boldsymbol{k}_{\perp}\right)\cdot\boldsymbol{\hat{z}} & 0\\
0 & -\left(\boldsymbol{\sigma}_{\perp}\times\boldsymbol{k}_{\perp}\right)\cdot\boldsymbol{\hat{z}}
\end{array}\right)
\end{equation}
 This effective 2D Hamiltonian can be reduced to two Weyl Hamiltonians
of the form $H_{2D}^{2\times2}=\pm\hbar v\left(\boldsymbol{\sigma}_{\perp}\times\boldsymbol{k}_{\perp}\right)\cdot\boldsymbol{\hat{z}}$.
It is important to note that both 2-spinors of $\chi_{\pm}$, the
2-spinor $\chi_{\pm}^{L^{-}}$ of the $L^{-}$ band and the 2-spinor
$\chi_{\pm}^{L^{+}}$ of the $L^{+}$ band have the same helicity,
in contrast to the commonly used Weyl Hamiltonians $H_{W}(\mathbf{k})=\pm\hbar v\mathbf{\mathbf{\mathbf{\boldsymbol{\sigma}\cdot}k}}$.
The reason for this is that the two 2-spinors are coupled through
the mass term $\Delta(z)$ in $z$-direction, as given in the 3D Hamiltonian
in Eq.~(\ref{eq:1}).

\section{Optical transition matrix elements}

\label{sec:matrix_elements} We calculate the low-energy transitions
around the $L$ valley that is lifted up along the $z$-direction
from the other three $L$ valleys. With the proper choice of uniform
strain, composition and layer width, there exist practically gapless
helical states for the {[}111{]} valley inside the gapped states of
the oblique valleys.\cite{key-23-1} In unstrained Pb$_{1-x}$Sn$_{x}$Te,
band inversion occurs simultaneously at four $L$ points and the phase
is topologically trivial. For most experiments, in a structure with
a layer of thickness d $\approx10$ nm between the two interfaces
in a PbTe/Pb$_{0.31}$Sn$_{0.69}$Te/PbTe heterostructure, dispersion
of the {[}111{]} valley states can be assumed to be gapless while
the states in the oblique valleys are gapped.\cite{key-23-1} The
interface can be modeled with the bulk of Pb$_{0.31}$Sn$_{0.69}$Te
and PbTe with bandgaps of, respectively, -0.187 and 0.187 eV, so that
Weyl fermions are generated at the two interfaces. Here, the bandgap
formula provided in Ref.~\onlinecite{Yusheng} was used. It is
to be noted that localized spin states of 2D Weyl fermions in 3D TI
are solutions of the $\mathbf{k}\cdot\mathbf{p}$ Hamiltonian given
in Eq.~(\ref{eq:1}).

Now we proceed to calculate the optical selection rules for the excitation
of electron-hole pairs, keeping in mind that the Dirac equation provides
an effective description of the two-band system consisting of the
$L^{\mp}$ bands. The $\mathbf{k}\cdot\mathbf{p}$ Hamiltonian contains
also a quadratic term in the momenta,\cite{Nimtz:1983} namely 
\begin{equation}
H_{q}=\left(\begin{array}{cc}
\frac{\left(p_{z}+eA_{z}\right)^{2}}{2m_{\Vert}^{-}}+\frac{\left(\mathbf{p}_{\bot}+e\mathbf{A}_{\bot}\right)^{2}}{2m_{\bot}^{-}} & 0\\
0 & \frac{\left(p_{z}+eA_{z}\right)^{2}}{2m_{\Vert}^{+}}+\frac{\left(\mathbf{p}_{\bot}+e\mathbf{A}_{\bot}\right)^{2}}{2m_{\bot}^{+}}
\end{array}\right),
\end{equation}
where $m_{\Vert}^{\mp}$ and $m_{\bot}^{\mp}$ are the longitudinal
and transverse effective masses of the $L^{\mp}$ bands, respectively.
Through minimal coupling the quadratic term leads to a linear term
in the momentum, which we need to take into account. Hence, in the
presence of electromagnetic radiation, the total Hamiltonian for the
Dirac particle is given by \begin{widetext} 
\begin{eqnarray}
H_{{\rm tot}} & = & v_{\Vert}\alpha_{z}\left(\hat{p}_{z}+eA_{z}\right)+v_{\bot}\boldsymbol{\alpha}_{\bot}\cdot\left(\mathbf{\hat{p}}+e\mathbf{A}_{\perp}\right)+\beta\Delta(z)+(e/m)\mathbf{A}\cdot\mathbf{p}\nonumber \\
 & = & \left(\begin{array}{cc}
\Delta(z)+e\left(\frac{p_{z}A_{z}}{m_{\parallel}^{-}}+\frac{\mathbf{p}_{\perp}\cdot\mathbf{A}_{\perp}}{m_{\perp}^{-}}\right) & v_{\Vert}\sigma_{z}\left(\hat{p}_{z}+eA_{z}\right)+v_{\bot}\mathbf{\mathbf{\mathbf{\boldsymbol{\sigma}_{\bot}}}}\cdot\left(\mathbf{\hat{p}}+e\mathbf{A}_{\perp}\right)\\
v_{\Vert}\sigma_{z}\left(\hat{p}_{z}+eA_{z}\right)+v_{\bot}\mathbf{\mathbf{\mathbf{\boldsymbol{\sigma}_{\bot}}}}\cdot\left(\mathbf{\hat{p}}+e\mathbf{A}_{\perp}\right) & -\Delta(z)+e\left(\frac{p_{z}A_{z}}{m_{\parallel}^{+}}+\frac{\mathbf{p}_{\perp}\cdot\mathbf{A}_{\perp}}{m_{\perp}^{+}}\right)
\end{array}\right).
\end{eqnarray}
\end{widetext} where $A=(A_{z},\;\boldsymbol{A}_{\perp})$ is the
vector potential, $\boldsymbol{\alpha}=(\alpha_{z},\:\boldsymbol{\alpha}_{\bot})$
and $\beta$ are the Dirac matrices $\boldsymbol{\alpha}_{i}=\left(\begin{array}{cc}
0 & \boldsymbol{\sigma}_{i}\\
\boldsymbol{\sigma}_{i} & 0
\end{array}\right)$, $\beta=\left(\begin{array}{cc}
I & 0\\
0 & -I
\end{array}\right)$, and $\mathbf{E}=\partial\mathbf{A}/\partial t$ in the Coulomb gauge.
We identify the interaction Hamiltonian as 
\begin{align}
 & H_{int}=ev_{\Vert}\alpha_{z}A_{z}+ev_{\bot}\boldsymbol{\alpha}_{\bot}\cdot\boldsymbol{A}_{\bot}+(e/m)\mathbf{A}\cdot\mathbf{p}\\
= & \left(\begin{array}{cc}
e\left(\frac{p_{z}A_{z}}{m_{\parallel}^{-}}+\frac{\mathbf{p}_{\perp}\cdot\mathbf{A}_{\perp}}{m_{\perp}^{-}}\right) & ev_{\Vert}\sigma_{z}A_{z}+ev_{\bot}\mathbf{\mathbf{\mathbf{\boldsymbol{\sigma}_{\bot}}}}\cdot\mathbf{A}_{\perp}\\
ev_{\Vert}\sigma_{z}A_{z}+ev_{\bot}\mathbf{\mathbf{\mathbf{\boldsymbol{\sigma}_{\bot}}}}\cdot\mathbf{A}_{\perp} & e\left(\frac{p_{z}A_{z}}{m_{\parallel}^{+}}+\frac{\mathbf{p}_{\perp}\cdot\mathbf{A}_{\perp}}{m_{\perp}^{+}}\right)
\end{array}\right).\nonumber 
\end{align}
It will turn out that only interband transitions contribute for a
2D interface, whereas both interband and intraband transitions contribute
in the case of a 3DTI quantum dot. It is important to note that $v_{\Vert}=P_{\Vert}/m_{0}$
and $v_{\bot}=P_{\bot}/m_{0}$ include the Kane interband matrix elements
$\mathbf{P}=\left\langle u_{\mathbf{k}_{\mathrm{f}}}^{\mp}\left|\mathbf{\hat{P}}\right|u_{\mathbf{k}_{\mathrm{I}}}^{\pm}\right\rangle $,
where $u_{\mathbf{k}}^{\mp}$ are the Bloch's functions for the $L^{\mp}$
bands. This means that the interband transitions are governed by the
interband Hamiltonian $H_{inter}=ev_{\Vert}\alpha_{z}A_{z}+ev_{\bot}\boldsymbol{\alpha}_{\bot}\cdot\boldsymbol{A}_{\bot}$,
where the Dirac $\boldsymbol{\alpha}$- matrices couple the $L^{-}$
band with the $L^{+}$ band. The Hamiltonian $H_{intra}=(e/m)\mathbf{A}\cdot\mathbf{p}$
accounts for intraband transitions with $\mathbf{\hat{p}}$ operating
on the envelope wavefunctions only. $H_{intra}$ is proportional to
the identity in 4-spinor space and therefore couples the $L^{-}$
band to itself and the $L^{+}$ band to itself. Thus the interband
Hamiltonian $H_{inter}$ and the intraband Hamiltonian $H_{intra}$
are not equivalent in this description. On the one hand, $H_{inter}$
gives rise to interband transitions because it contains the Kane interband
matrix elements $P_{\bot}$ and $P_{\Vert}$. On the other hand, $H_{intra}$
gives rise to intraband transitions because the term $(e/m)\mathbf{A}\cdot\mathbf{p}$
operates on the envelope wavefunctions. 

We start with calculating the interband matrix elements which are
given by the off diagonal elements of the interaction Hamiltonian.
We identify $j_{z}=ev_{\Vert}\Psi^{\dagger}\alpha_{z}\Psi$ and $\mathbf{j}_{\bot}=ev_{\bot}\Psi^{\dagger}\boldsymbol{\alpha}_{\bot}\Psi$
as the longitudinal and transverse relativistic current densities,
respectively. Therefore, the evaluation of the optical transition
matrix elements is reduced to calculating the matrix elements of $\alpha_{i}$. 

\begin{figure}
\includegraphics[width=5cm]{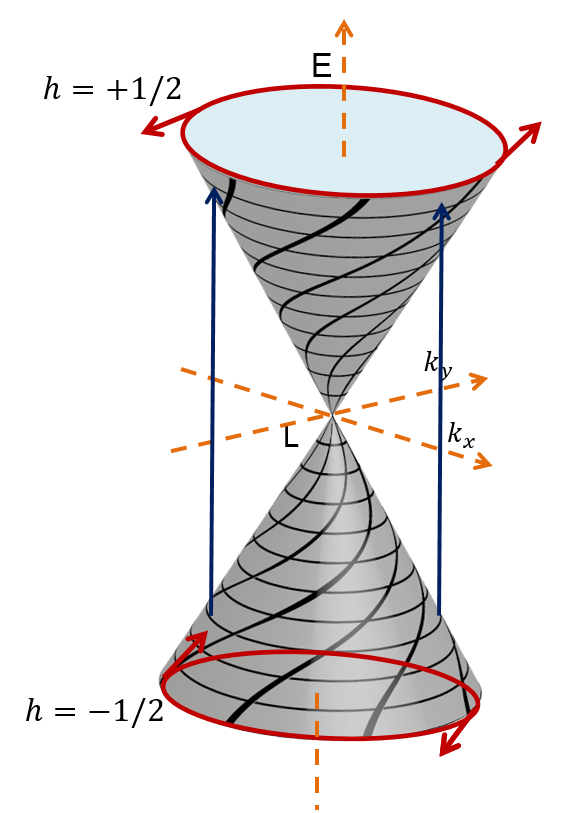}\caption{The interband transitions for the spin selection rules in 3D TIs.
The Dirac cone represents the component of the Weyl states. The interband
transitions occur between positive and negative energy solutions.
The helicity of the band is represented by $h=+1/2$ $\left(h=-1/2\right)$
for the positive energy solution (negative energy solution)\textcolor{red}{.
\label{Cones_transitions}}}
\end{figure}

The optical transition matrix elements involve the integral over the
envelope functions and the periodic part of the Bloch functions. The
integral over the envelope function can be carefully separated out
from the remaining part, similarly to the case of wide-bandgap semiconductor
materials.\cite{Yu&Cardona} The idea is to separate the slowly varying
envelope part from the rapidly varying periodic part of the total
wavefunction. For that we need to first replace the position vector
by ${\bf {r}}={\bf {r'}}+{{\bf {R}}_{m}}$, where ${\bf {R}}_{m}$
is a lattice vector and ${\bf {r'}}$ is a vector within one unit
cell. Writing the vector potential $\mathbf{A}=(A_{x0},A_{y0},A_{z0})e^{i\mathbf{q}\cdot\mathbf{r}}$
and taking advantage of the periodicity $u_{\mathbf{k}}^{L^{\pm}}({\bf {r'}}+{{\bf {R}}_{m}})=u_{\mathbf{k}}^{L^{\pm}}({\bf {r'}})$
and the fact that $\phi_{\pm}^{L^{\pm}}({\bf {r'}}+{{\bf {R}}_{m}})\approx\phi_{\pm}^{L^{\pm}}({{\bf {R}}_{m}})$,
we obtain 
\begin{eqnarray}
\left\langle \Phi_{f}\left|H_{{\rm int}}\right|\Phi_{I}\right\rangle  & \approx & \frac{e}{m_{0}}\sum\limits _{m}\left|F(z)\right|^{2}{e^{i\left({{\bf {q}}+{{\bf {k}}_{\mathrm{I}}}-{{\bf {k}}_{\mathrm{f}}}}\right)\cdot{{\bf {R}}_{m}}}}\nonumber \\
 &  & \times\sum_{i}A_{i0}\int\limits _{\Omega}u{{_{{{\bf {k}}_{\mathrm{f}}}}^{{L^{-}}}}^{*}}({\bf {r'}})\hat{P}_{i}u_{{{\bf {k}}_{\mathrm{I}}}}^{{L^{+}}}({\bf {r'}})\nonumber \\
 &  & \times{e^{i\left({{\bf {q}}+{{\bf {k}}_{\mathrm{I}}}-{{\bf {k}}_{\mathrm{f}}}}\right)\cdot{\bf {r'}}}}{d^{3}}r'\nonumber \\
 &  & \times\left(\chi_{\mathrm{f}}^{{L^{-}}}{\sigma_{i}}\chi_{I}^{{L^{+}}}+\chi_{f}^{{L^{+}}}{\sigma_{i}}\chi_{I}^{{L^{-}}}\right)
\end{eqnarray}
for an optical transition from the initial state $\left|\Phi_{I}\right\rangle $
to the final state $\left|\Phi_{f}\right\rangle $. $\Omega$ is the
volume of the unit cell. By applying the secular approximation to
the term with the exponential function ${e^{i\left({{\bf {q}}+{{\bf {k}}_{\mathrm{I}}}-{{\bf {k}}_{\mathrm{f}}}}\right)\cdot{{\bf {R}}_{m}}}}$,
we obtain ${{\bf {k}}_{\mathrm{f}}}={\bf {q}}+{{\bf {k}}_{\mathrm{I}}}$,
which ensures momentum conservation in the plane of the interface.
Using the normalization $\int_{-\infty}^{\infty}\left|F(z)\right|^{2}dz=1$,
the optical matrix element is well approximated by 
\begin{eqnarray}
\left\langle \Phi_{f}\left|H_{{\rm int}}\right|\Phi_{I}\right\rangle  & \approx & \frac{e}{m_{0}}\sum_{i}A_{i0}\int\limits _{\Omega}u{{_{{{\bf {k}}_{\mathrm{f}}}}^{{L^{-}}}}^{*}}({\bf {r'}})\hat{P}_{i}u_{{{\bf {k}}_{\mathrm{I}}}}^{{L^{+}}}({\bf {r'}}){d^{3}}r'\nonumber \\
 &  & \times\left(\chi_{\mathrm{f}}^{{L^{-}}}{\sigma_{i}}\chi_{I}^{{L^{+}}}+\chi_{f}^{{L^{+}}}{\sigma_{i}}\chi_{I}^{{L^{-}}}\right)\nonumber \\
 & = & eA_{z0}v_{\Vert}\left\langle \chi_{f}\left|\alpha_{z}\right|\chi_{I}\right\rangle \nonumber \\
 &  & +eA_{x0}v_{\bot}\left\langle \chi_{f}\left|\alpha_{x}\right|\chi_{I}\right\rangle \nonumber \\
 &  & +eA_{y0}v_{\bot}\left\langle \chi_{f}\left|\alpha_{y}\right|\chi_{I}\right\rangle .
\end{eqnarray}
Note that in contrast to semiconductor quantum wells where the overlap
between electron and hole envelope wavefunctions is smaller than 1
in general, here the overlap between Weyl envelope wavefunctions is
$\int_{-\infty}^{\infty}\left|F(z)\right|^{2}dz=1$. We assume that
the wavelength of incoming photon is small compared to the lattice
constant. This means we can use the dipole approximation: $\mathbf{A}\thickapprox(A_{x0},A_{y0},A_{z0})$.
Since there is no net momentum transfer the directions of the initial
and final momentum vectors are the same; i.e. we consider only vertical
transitions. For the $\alpha$ matrix elements we obtain the following
interband matrix elements: 
\begin{equation}
\left\langle \chi_{+}\left|\alpha_{x}\right|\chi_{-}\right\rangle =4i\sin\theta\qquad\left\langle \chi_{+}\left|\alpha_{y}\right|\chi_{-}\right\rangle =-4i\cos\theta,\label{eq:15}
\end{equation}
These transitions are vertical. The z-component of the matrix element
of $\boldsymbol{\alpha}$ vanishes. 

The Kane interband matrix element can be calculated explicitly. The
periodic function $u_{\mathbf{k}}(\mathbf{r})$ can be written as
$u_{\mathbf{k}}^{L^{\pm}}=\underset{\mathbf{G}}{\sum}a_{L^{\pm}}(\mathbf{G})e^{i\mathbf{G}\cdot\mathbf{r}}$,
where $\mathbf{G}$ is the reciprocal lattice vector and $a_{L^{\pm}}(\mathbf{G})$
are the expansion cofficients for the $L^{\pm}$ bands. The Kane interband
matrix elements can be evaluated as 
\begin{align}
\underset{\Omega}{\int}u_{\mathbf{k}_{\mathrm{f}}}^{L^{-}*}\mathbf{\hat{P}}u_{\mathbf{k}_{\mathrm{I}}}^{L^{+}}d^{3}r=\underset{\mathbf{G}_{\mathrm{f}},\mathbf{G}_{\mathrm{I}}}{\sum}\underset{\Omega}{\int}e^{-i\left(\mathbf{G}_{\mathrm{f}}-\mathbf{G}_{\mathrm{I}}\right)\cdot\mathbf{r}}d^{3}r\nonumber \\
\times a_{L^{-}}^{*}\left(\mathbf{G_{\mathrm{f}}}\right)\mathbf{G}_{\mathrm{I}}a_{L^{+}}\left(\mathbf{G_{\mathrm{I}}}\right)
\end{align}
For the vertical transitions $k_{f}\thickapprox k_{I}$ and $\underset{\Omega}{\int}e^{-i\left(\mathbf{G}_{\mathrm{f}}-\mathbf{G}_{\mathrm{I}}\right)\cdot\mathbf{r}}d^{3}r=\delta\left(\mathbf{G}_{\mathrm{f}}-\mathbf{G}_{\mathrm{I}}\right)$,
we obtain 
\begin{equation}
\underset{\Omega}{\int}u_{\mathbf{k}}^{L^{-}*}\mathbf{\hat{P}}u_{\mathbf{k}}^{L^{+}}d^{3}r=\underset{\mathbf{G}}{\sum}\mathit{a_{L^{-}}^{*}}(\mathbf{G})\mathbf{G}\mathit{a}_{L^{+}}(\mathbf{G}).\label{eq:16}
\end{equation}

\begin{figure}
\includegraphics[width=8.5cm]{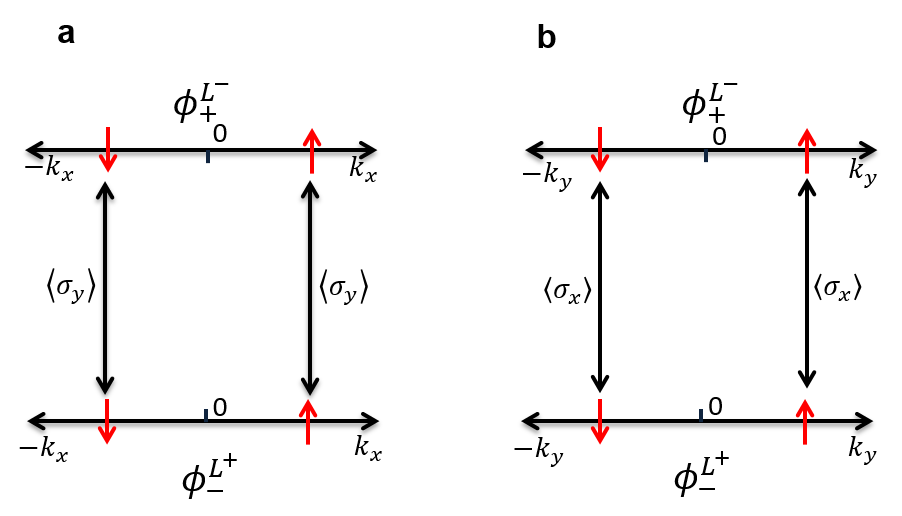}\caption{Spin selection rules in 3D TIs. The states are labeled with $\phi_{\pm}^{L^{\pm}}$.
The transitions are vertical conserving the spin's direction. The
direction of the momentum is shown along x-axis (a) in which case
the polarization of the light couples the spin pointing along y-axis
and along y-axis (b) in which case the polarization of the light couples
the spin pointing along x-axis. In each case the spin points perpendicular
to the momentum (see Fig.~\ref{Cones_transitions}). \label{SpinSelectionRules}}
\end{figure}

The diagonal matrix elements of the interaction Hamiltonian give rise
to the intraband transitions. As stated above, the intraband matrix
elements operate on the envelope functions only and thus couple to
the $L^{-}$ band to itself and $L^{+}$ band to itself. In the electric
dipole approximation the transitions within the same energy solutions
are absent. The intraband matrix elements for the transitions occurring
between the positive and negative energy solutions are given by
\begin{eqnarray}
\left\langle \Phi_{+}\left|\hat{\mathbf{e}}.\mathbf{p}\right|\Phi_{-}\right\rangle & = & \left[\left\langle \chi_{+}^{L^{-}}\left|\chi_{-}^{L^{-}}\right\rangle +\right.\left\langle \chi_{+}^{L^{+}}\left|\chi_{-}^{L^{+}}\right\rangle \right.\right]
\nonumber\\
& & \times \left\langle F(\mathbf{r})\left|\hat{\mathbf{e}}.\mathbf{p}\right|F(\mathbf{r})\right\rangle ,
\end{eqnarray}
where the Bloch's functions are already integrated to unity. From
the Eqs.~(\ref{eq:6}) and (\ref{eq:7}), it is seen that the 2-component
spinors for the same band corresponding to different energy solutions
are orthogonal to each other: $\left\langle \chi_{+}^{L^{-}}\left|\chi_{-}^{L^{-}}\right\rangle \right.=0$
and $\left\langle \chi_{+}^{L^{+}}\left|\chi_{-}^{L^{+}}\right\rangle \right.=0$.
This implies that $\left\langle \Phi_{+}\left|\hat{\mathbf{e}}.\mathbf{p}\right|\Phi_{-}\right\rangle =0$.
This is, indeed, different from the case of wide bandgap semiconductor
materials where we usually have both intraband and interband transitions. 

\label{sec:selection_rules} In Fig.~\ref{SpinSelectionRules} we
show the possible transitions allowed by the spin selection rules.
In each case transitions happen between $L^{+}$ and $L^{-}$ band
each band between positive energy solution and negative energy solution.
Since we use the dipole approximation initial and final momentum point
in same direction and have the same magnitude; i.e. the transitions
are vertical. If the momentum vector in one of the bands points along
the $x$-axis, according the Eq.~(\ref{eq:15}), the polarization of
the photon couples to the spin pointing along $y$-axis. If the momentum
vector in one of the bands points along the $y$-axis, the polarization
of the photon couples to the spin pointing along $x$-axis. In each
case the spin's direction is conserved.

\section{Bandstructure calculation}

\label{sec:bandstructure} 
\begin{figure}
\includegraphics[width=8.5cm]{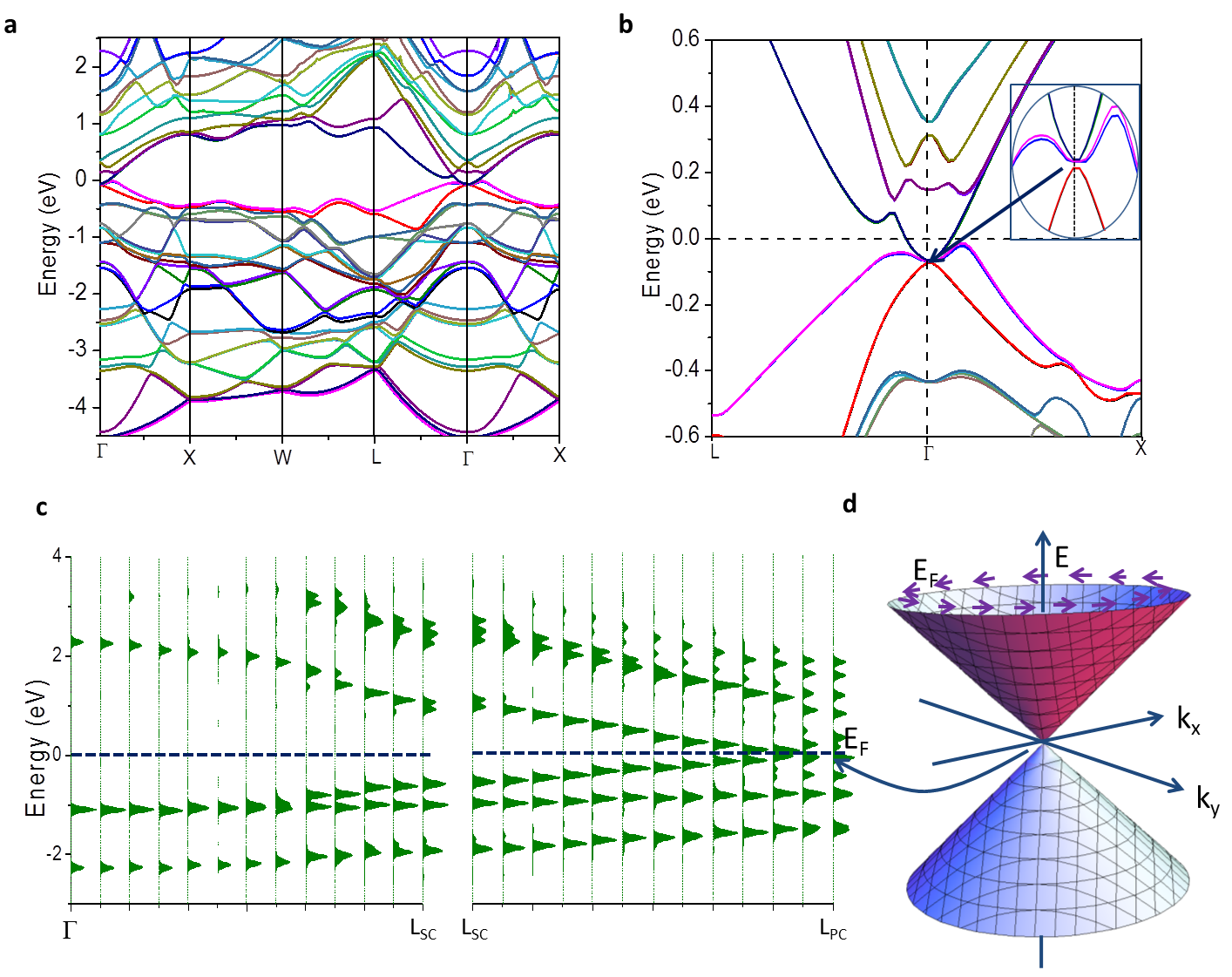}\caption{Bulk bandstructure of Pb$_{1-x}$Sn$_{x}$Te at $x=0.375$ doping
level including spin-orbit coupling. The crossing has been reported
around $x=0.35$.\cite{key-12} In the supercell Brillouin zone bands
are folded from the neighboring Brillouin zones into the first Brillouin
zone {[}(a) and (b){]}. A small band gap of 1.8 meV appears at the
$\Gamma$ point of the supercell Brillouin zone, which corresponds
to the band gap minimum at the $L$ point in the unfolded primitive
cell Brillouin zone, as shown in \textbf{c}. $L_{SC}$ and $L_{PC}$
are $L$ points in the supercell Brillouin zone and the primitive
cell Brillouin zone, respectively. The solid green color peaks in
(c) denotes the spectral functions.\cite{key-18} Bands of opposite
parity nearly cross at around 67 meV below the Fermi level at the
$L$ point where a single Dirac point is observed (d). \label{BandStructure}}
\end{figure}

In order to know the relative strength
of the transitions, it is important to calculate the complete bandstructures
of Pb$_{1-x}$Sn$_{x}$Te, which also provides the cofficients of
the periodic part of Bloch functions that appear in the selection
rules. Fig.~\ref{BandStructure} shows the calculations of the complete
bulk bandstructures of Pb$_{1-x}$Sn$_{x}$Te at 37.5\% doping by
Sn impurities in a supercell Brillouin zone using density functional
theory within PAW approximation as implemented in VASP.\cite{key-24,key-25,key-26}
We unfold the bandstructures along the $\Gamma$ to $L$ point in
the first Brillouin zone using unfolding recipes.\cite{key-18}

The unfolded bandstructure is equivalent to the folded bandstructure
in terms of the magnitude of band separation as required by the energy
conservation law. The point $L_{PC}$ in the unfolded bandstructure
is a mirror image of the point $\Gamma$ in the folded bandstructure,
therefore, bands appear with the same dispersion as they were before
unfolding. In the unfolded bandstructures, bands around the $L$ point
are almost linear, which is best described by Weyl fermions. The Dirac
point appears at 67 meV below the Fermi level at the $L$ point. The
valence band maximum is derived from the p orbitals of Pb and Sn hybridized
with the s orbital of Te and the conduction band minimum is derived
from the s orbitals of Pb and Sn hybridized with the p orbital of
Te. They have opposite parity, thus making interband transitions allowed.
As measured in the experiment, the anisotropy in the crystal structure
gives velocity components as $v_{\bot}=4.2\times10^{5}$ m/s and $v_{\Vert}=1.7\times10^{5}$
m/s.\cite{Xu} 

The localization length $l{}_{o}$ for the Weyl states along $z$-axis
can be obtained using our calculated band gap of 350 meV including
spin-orbit coupling for PbTe. Using the band velocity, $v_{\parallel}=1.7\times10^{5}$
m/s, we obtain $l{}_{o}=0.32$ nm. This length measures the characteristic
scale of the confinement of Weyl states along $z$-axis at the interface.

\section{\label{Faraday-Effect for 3D TIs}Faraday Effect for 3D TIs }

In Refs.~\onlinecite{Leuenberger:2005,Leuenberger:2006,Seigneur:2011,Gonzalez:2010,Seigneur:2010}
it has been shown that the single-photon Faraday rotation can be used
for quantum spin memory and quantum teleportation and quantum computing
with wide-bandgap semiconductor QDs. The conditional Faraday rotation
can be used for optical switching of classical information\cite{Thompson:2009}.
A single-photon Mach-Zehnder interferometer for quantum networks based
on the single-photon Faraday effect has been proposed in Ref.~\onlinecite{Seigneur:2008}.
In Ref. \onlinecite{Berezovsky} a single spin in a wide-bandgap
semiconductor QD was detected using the Faraday rotation. It is evident
from the calculation above that we have strict optical selection rules
for the $x$ and $y$ polarization states of the photons. We show
below that these strict optical selection rules give rise to a giant
Faraday effect due to Pauli exclusion principle for 3D TIs using our
continuum eigenstates. 

\begin{figure}
\includegraphics[width=8.5cm]{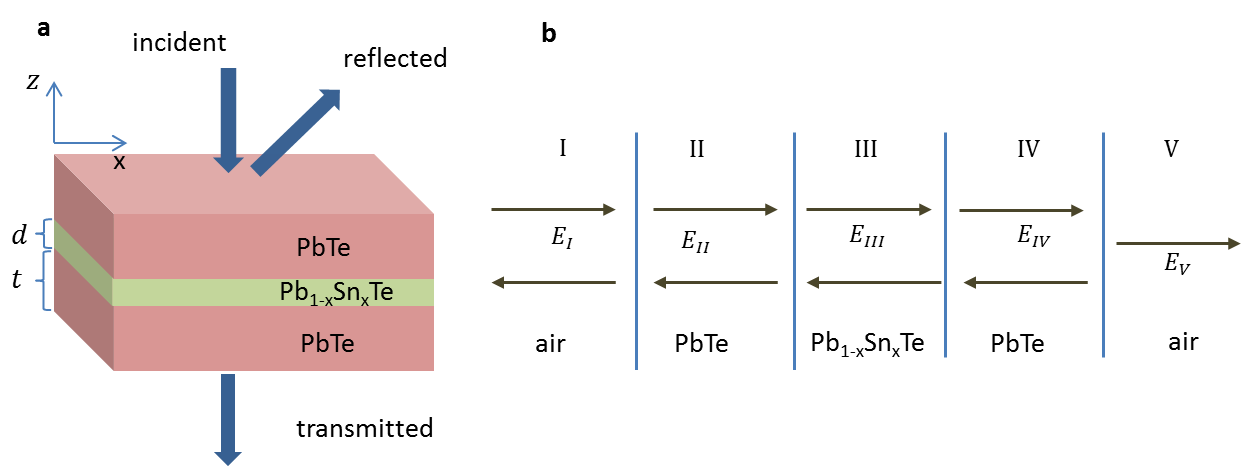}\caption{\textbf{a}. A slab of thickness $d=10$ nm of 3D TI material Pb$_{1-x}$Sn$_{x}$Te
is sandwiched by PbTe with thickness t. This structure can have Weyl
fermions at the interface with zero bandgap at one of the L point
in the Brillouin zone while the rest of the L points have non-zero
bandgaps due to the interactions between the L valleys of the two
interfaces.\cite{key-23-1} \textbf{b}. Solutions inside and outside
the material can be found by dividing the geometry into five different
regions, $I$, $II$, $III$, $IV$ and $IV$ with the fields $E_{I}$,
$E_{II}$, $E_{III}$, $E_{IV}$ and $E_{V}$, \label{fig:layers}}
\label{3DTI_slab}
\end{figure}

Let us consider the PbTe/Pb$_{0.31}$Sn$_{0.69}$Te/PbTe heterostructure
shown in Fig. \ref{fig:layers}. A laser pump beam excites e-h pairs
at the two interfaces between Pb$_{0.31}$Sn$_{0.69}$Te and PbTe.
It is important to understand the working scheme of the driving fields
and a dynamics of the hot carriers in the excited states so that maximum
Faraday effect can be achieved in an experiment. The e-h pairs pumped
by the driving field relax mainly through the electron-phonon interaction
before they recombine. On a time scale of several hundred ps, the
electrons and holes cool down after the driving field is turned off\cite{Zhang&Wu}.
Due to the presence of the strong spin-orbit coupling in 3D TI, the
induced spin polarization relaxes on a time scale of the momentum
scattering. As calculated in Ref. \cite{Zhang&Wu}, the spin polarization
decays rapidly within a time of the order of $T_{2}=$0.01\textendash{}0.1
ps, which results in a loss of spin coherence. Consequently, it is
very difficult to measure the Faraday effect after the pump pulse
is turned off. To circumvent the problem of fast spin decoherence,
we suggest to use both the pump and the probe fields simultaneously,
thereby maintaining the coherence of the induced spin polarization
in the excited states. Therefore, the probe field experiences a response
from the spin polarized carriers. We use an off-resonant probe field
with detuning energy of around 10 meV.

Let us write the light-matter interaction Hamiltonian as $H_{int}=ev\boldsymbol{\alpha}\cdot\boldsymbol{A}$,
which contains the interband term only because the intaband term is
zero, as shown in Sec. \ref{sec:matrix_elements}. Without loss of
generality, the anisotropy coming from the band velocity can be introduced
back into the solutions at a later time. Since the incident light
is a plane wave with wave vector $\mathbf{q}$ and frequency $\omega$
and the electric field component is $E=-\partial\boldsymbol{A}/\partial t$,
the interaction Hamiltonian reads 
\begin{equation}
H_{int}=\frac{ePE_{0}}{im_{0}\omega}\left(e^{i\left(\mathbf{q\cdot r}-\omega t\right)}-e^{-i\left(\mathbf{q\cdot r}-\omega t\right)}\right)\boldsymbol{e\cdot\alpha}
\end{equation}
where $P=m_{0}v$ is the Kane interband matrix element. The transition
rate for a single can be calculated using Fermi's golden rule, 
\begin{eqnarray}
W_{fI} & = & \frac{2\pi}{\hbar}\left(\frac{eE_{0}P}{m_{0}\omega}\right)^{2}\left|\left\langle \Phi_{f}\right|\boldsymbol{e\cdot\alpha}\left|\Phi_{I}\right\rangle \right|^{2}\nonumber \\
 &  & \times n_{I}(t)\left[1-n_{f}(t)\right]\delta\left(\varepsilon_{f}-\varepsilon_{I}\mp\hbar\omega\right)
\end{eqnarray}
where $n_{I,f}$ is the population distribution function for the initial
and final states, $\varepsilon_{F}$ is the Fermi energy, $\left|\Phi_{I}\right\rangle $
denotes the initial Weyl state, $\left|\Phi_{f}\right\rangle $ denotes
the final Weyl state, and the - sign in front of $\hbar\omega$ corresponds
to absorption and the + sign to emission. Thus, the absorption of
energy per spin state is $\mathcal{P}=\hbar\omega\sum_{I,f}W_{fI}$.
Comparing with the total power $\mathcal{P}=2\sigma_{1}VE_{0}^{2}$
dissipated in the system area $V$, where $\sigma=\sigma_{1}+i\sigma_{2}$
is the complex conductivity, and including absorption and emission,
it follows that the real part of the conductivity is 
\begin{eqnarray}
\sigma_{1} & = & \frac{\pi e^{2}P^{2}}{Vm_{0}^{2}\omega}\sum_{I,f}\left|\left\langle \Phi_{f}\right|\boldsymbol{e\cdot\alpha}\left|\Phi_{I}\right\rangle \right|^{2}\nonumber \\
 &  & \times\left[n_{I}(t)-n_{f}(t)\right]\delta\left(\varepsilon_{f}-\varepsilon_{I}-\hbar\omega\right)
\end{eqnarray}
which can be written in terms of the oscillator strengths $f_{fI}=\left(\frac{2P^{2}}{\hbar m_{0}\omega_{fI}}\right)\left|\left\langle \Phi_{f}\right|\boldsymbol{e\cdot\alpha}\left|\Phi_{I}\right\rangle \right|^{2}$,
\begin{equation}
\sigma_{1}\left(\omega\right)=\frac{\pi e^{2}}{2m_{0}V}\sum_{fI}f_{fI}\left[n_{I}(t)-n_{f}(t)\right]\delta\left(\varepsilon_{f}-\varepsilon_{I}-\hbar\omega\right)\label{eq:21}
\end{equation}
Using the Kramers-Kronig relations $\sigma_{2}\left(\omega\right)$
can be obtained. It is important to note that $\sigma_{1}\left(\omega\right)$
is equivalent to the imaginary part of the dieletric function, $\epsilon\left(\omega\right)$.
The physical significance of $\sigma_{1}\left(\omega\right)$ and
$\sigma_{2}\left(\omega\right)$ appear in different way, $\sigma_{1}\left(\omega\right)$
being for the dissipiation while $\sigma_{2}\left(\omega\right)$
for the polarization. 

We calculate now the Faraday rotation angle due to Pauli exclusion
principle between the initial and final continuum states. In order
to this, a strong $\pi$-pulse of the laser pump beam is used to excite
e-h pairs. The direction of the polarization can be along $x-$and
$y-$axis. The dynamics of the excitation of e-h pairs can be described
by the optical Bloch equations \cite{Haug&Koch}. Due to the large
screening the exciton binding energies in perpendicular and parallel
directions are small, i.e. $E_{b\bot}=143$ $\mu$eV and $E_{b\Vert}=1.68$
meV.\cite{Murphy:2006} Therefore, we can safely neglect the Coulomb
interaction. Then the time dependences of the polarization $P_{\mathbf{k}}$
and the electron population distribution $n_{e,\mathbf{k}}$ for the
state $\mathbf{k}$ are given by
\begin{eqnarray}
\frac{dP_{\mathbf{k}}}{dt} & = & i\varepsilon_{g}P_{\mathbf{k}}+i(n_{e,\mathbf{k}}+n_{h,\mathbf{k}}-1)\omega_{R,\mathbf{k}}, \label{eq:22}
\\
\frac{dn_{e,\mathbf{k}}}{dt} & = & -2Im(\omega_{R,\mathbf{k}}P_{\mathbf{k}}^{*}), \label{eq:23}
\end{eqnarray}
where $\varepsilon_{g}=\varepsilon_{e,\mathbf{k}}+\varepsilon_{h,\mathbf{k}}$
$\varepsilon_{e,\mathbf{k}}$ and $\varepsilon_{h,\mathbf{k}}$ are
the electron and hole kinetic energies, respectively, in the state
$\mathbf{k}$, and $\omega_{R,\mathbf{k}}$ is the Rabi frequency.
An equation similar to Eq.~(\ref{eq:23}) can be written for the hole
distribution function $n_{h,\mathbf{k}}$. It is to be noted that
$n_{h,\mathbf{k}}=n_{e,\mathbf{k}}$. In the rotating frame approximation,
$P_{\mathbf{k}}(t)=\tilde{P}(t)e^{-i\varepsilon_{g}t}$ and $\omega_{R,\mathbf{k}}(t)=\omega_{o,\mathbf{k}}e^{-i\varepsilon_{g}t}$.
Using this Eqs.~(\ref{eq:22}) and (\ref{eq:23}) yield $d\tilde{\eta}_{\mathbf{k}}/dt=2(n_{e,\mathbf{k}}-1)\omega_{o,\mathbf{k}}$
and $\tilde{n}_{e,\mathbf{k}}=-2\omega_{o}\tilde{\eta}$, where $\tilde{\eta}=(\tilde{P}-\tilde{P}^{*})/2i$.
These two equations can be solved for $n_{e,\mathbf{k}}$. We obtain,
$n_{e,\mathbf{k}}=\frac{1}{2}\left[1-\mathrm{\cos}(2\omega_{o,\mathbf{k}}t)\right]$.
A similar solution can be obtained for $n_{h,\mathbf{k}}$. For $2\omega_{o,\mathbf{k}}t=m\pi$, 
$n_{e,\mathbf{k}}=1$ if $m$ is an odd integer, $n_{e,\mathbf{k}}=0$ if $m$ is an even integer, and $n_{e,\mathbf{k}}=1/2$ if 
$m$ is an odd half-integer. A strong $\pi$-pulse
excites the maximum number of electrons so that $n_{h,\mathbf{k}}\thickapprox1$
with $2\omega_{o,\mathbf{k}}t\thickapprox\pi$. In the absence of
Coulomb interaction the Rabi frequency can be written as $\omega_{o,\mathbf{k}}=d_{fI}\mathcal{E}\mathrm{\cos}\theta/\hbar$,
where $d_{fI}$ is a transitions dipole moment, $\mathcal{E}$ is
the strength of the electric field and $\theta$ is the direction
of polarization. 

It is useful to estimate the value of the Rabi frequency. The amplitude
of the electric field can be calculated as $\left|E_{S}\right|=\sqrt{2\mathit{\mathcal{S}}n/A\epsilon_{o}c}$,
where $\mathcal{S}$ is the power of the laser, $n$ is the index
of refraction of the medium through which the light propagates and
$A$ is the area of the aperture of the laser source. A laser power
of 0.5 mW with an area of the aperture of $10\:\mu m^{2}$ in a medium
with $n=5.8$ (for Pb$_{0.68}$Sn$_{0.32}$Te at room temperature)
can produce an electric field of $4.67\times10^{5}$ V/m. Using $v_{\perp}=4.2\times10^{5}$
m/s and the matrix elements from Eq.~(\ref{eq:15}), we obtain a maximum
Rabi frequency of $\omega_{o,max}=5.89\times10^{12}$/s which
occurs for $\theta=0$ or $\pi$. During the pump beam a laser probe
beam is incident on the double interface within the transparency region.
The polarization of this probe beam experiences the Faraday rotation
that we compute in the following.

The time dependence of the population becomes, $n_{e,\mathbf{k}}=\frac{1}{2}\left[1-\mathrm{\cos}\left(\frac{2d_{fI}\mathcal{E}\mathrm{\cos}\theta t}{\hbar}\right)\right]$.
The pump pulse duration, $T_{p}$, can be calculated using as $T_{p}=\pi\frac{\hbar}{2d_{fI}\mathcal{E}}$
. Probe and pump pulses are illuminated simultaneously to circumvent
the problem of decoherence of spin polarization, as described above.
Therefore, the probe pulse experiences the response from the average
spin coherent population distribution excited by the pump pulse. If
the probe pulse has the duration of $T_{r}=T_{p}$, the average population
distribution is calculated as, $\bar{n}_{e,\mathbf{k}}=\frac{1}{T_{p}}\underset{0}{\intop^{T_{p}}}n_{e,\mathbf{k}}\; dt$
which gives $\bar{n}_{e,\mathbf{k}}=\frac{1}{2}-\frac{1}{2}\frac{1}{\pi\mathrm{\cos}\theta}\mathrm{sin}(\pi\mathrm{\cos}\theta)$.
Since, $n_{v,\mathbf{k}}-n_{c,\mathbf{k}}=1-2n_{e,\mathbf{k}}$, for
$n_{v,\mathbf{k}}=n_{I}(t)$ and $n_{c,\mathbf{k}}=n_{f}(t)$, the
average of the net population distribution is $\overline{n_{I}(t)-n_{f}(t)}=\frac{1}{\pi\mathrm{\cos}\theta}\mathrm{sin}(\pi\mathrm{\cos}\theta)$.
If the probe pulse has the duration of $T_{r}=T_{p}/10$ and lasts
from the time $0.9T_{p}$ to the time $T_{p}$ of the pump pulse,
the average population distribution is $\bar{n}_{e,\mathbf{k}}=\frac{1}{T_{r}}\underset{0.9T_{p}}{\intop^{T_{p}}}n_{e,\mathbf{k}}\; dt$,
which gives $\bar{n}_{e,\mathbf{k}}=\frac{1}{2}-\frac{1}{2}\frac{10}{\pi\mathrm{\cos}\theta}\mathrm{sin}(\pi\mathrm{\cos}\theta)+\frac{1}{2}\frac{10}{\pi\mathrm{\cos}\theta}\mathrm{sin}(0.9\pi\mathrm{\cos}\theta)$.
Thus, we obtain $\overline{n_{I}(t)-n_{f}(t)}=\frac{10}{\pi\mathrm{\cos}\theta}\left[\mathrm{sin}(\pi\mathrm{\cos}\theta)-\mathrm{sin}(0.9\pi\mathrm{\cos}\theta)\right]$.
These average populations give rise to the Faraday rotation of the
probe field polarization. 

Now we proceed to describe the Faraday effect due to the 2D Weyl fermions
living at the interface of the 3D topological insulators. The difference
in the phase accumulated for the $x$ and $y$ polarization of the
light as it passes through the material is measured by the Faraday
rotation angle, which is solely due to the difference in response
of surface carriers to the $x$ and $y$ polarized light. This response
of the surface carriers at the two interfaces between Pb$_{0.31}$Sn$_{0.69}$Te
and PbTe is given by the optical conductivity tensor $\sigma_{ij}$,
$i=x,\, y$, $j=x,\, y$, which can be calculated by means of Eq.~(\ref{eq:21}). 
The interband matrix element, $\left|\left\langle \Phi_{f}\right|\boldsymbol{e\cdot\alpha}\left|\Phi_{I}\right\rangle \right|^{2}$,
for the linear polarization of light in $x$ and $y$ direction can
be written as 
\begin{eqnarray}
\left|\left\langle \Phi_{f}\right|\boldsymbol{e\cdot\alpha}\left|\Phi_{I}\right\rangle \right|^{2} & = & \left[\left|\left\langle \Phi_{f}\right|\alpha_{x}\left|\Phi_{I}\right\rangle \right|^{2}\right. \nonumber\\
& & +2\left\langle \Phi_{I}\right|\alpha_{x}\left|\Phi_{f}\right\rangle \left\langle \Phi_{f}\right|\alpha_{y}\left|\Phi_{I}\right\rangle
\nonumber\\ 
& & +\left.\left|\left\langle \Phi_{f}\right|\alpha_{y}\left|\Phi_{I}\right\rangle \right|^{2}\right].\label{eq:24}
\end{eqnarray}
The first and last terms of the RHS in Eq.~(\ref{eq:24}) are the matrix
elements that give rise to $\sigma_{xx}$ and $\sigma_{yy}$, respectively,
in $x$ and $y$ directions. The middle term gives rise to$\sigma_{xy}$.
Using Eq.~(\ref{eq:24}) and the average population distribution $\overline{n_{I}(t)-n_{f}(t)}$
after pumping using a linearly polarized light in $x$ direction in
Eq.~(\ref{eq:21}), one can solve for $\sigma_{xx}$ , $\sigma_{yy}$
and $\sigma_{xy}$. The summation can be changed into the integration
over the momentum space area, $\underset{fI}{\sum}\longrightarrow\left[1/\Omega_{\mathbf{k}}\left(2\pi\right)^{2}\right]\int k\, dk\int d\theta,\:0\leq\theta\leq2\pi$,
where $\Omega_{\mathbf{k}}$ is the cross sectional area of the Brillouin
zone. Using $d\varepsilon=\hbar v_{F}dk$, k-space integration can
be written as $\int k\, dk=\left[1/\hbar^{2}v_{F}^{2}\right]\int\varepsilon d\varepsilon$,
where $v_{F}=P/m_{0}$ is the Fermi velocity. As discussed above,
here we calculate the conductivity tensors for two examples of pulse
duration: $T_{r}=T_{p}$ and $T_{r}=T_{p}/10$. We obtain that $\sigma_{1xy}=0$.
This signifies that there is no transverse Hall effect with this type
of population distribution. If the polarization of the pump pulse
is in $y$ direction the transverse conductivity is still zero. $\sigma_{1xx}\left(\omega\right)$
and $\sigma_{1yy}\left(\omega\right)$ are calculated as follows:
Using the population distribution $\overline{n_{I}(t)-n_{f}(t)}$
in Eq.~(\ref{eq:22}), we obtain 
\begin{eqnarray}
\sigma_{1xx}\left(\omega\right)  & = & \frac{16\pi e^{2}}{\omega_{fI}\hbar^{2}\left(2\pi\right)^{2}}\underset{2\left|E_{F}\right|}{\overset{\infty}{\int}}\underset{0}{\overset{2\pi}{\int}}\varepsilon d\varepsilon\delta\left(\varepsilon_{f}-\varepsilon_{I}-\hbar\omega\right) \nonumber\\
& & \times\overline{n_{I}(t)-n_{f}(t)}\sin^{2}\theta d\theta, \label{eq:25} 
\\
\sigma_{1yy}\left(\omega\right) & = & \frac{16\pi e^{2}}{\omega_{fI}\hbar^{2}\left(2\pi\right)^{2}}\underset{2\left|E_{F}\right|}{\overset{\infty}{\int}}\underset{0}{\overset{2\pi}{\int}}\varepsilon d\varepsilon\delta\left(\varepsilon_{f}-\varepsilon_{I}-\hbar\omega\right) \nonumber\\
& & \times\overline{n_{I}(t)-n_{f}(t)}\cos^{2}\theta d\theta.\label{eq:26}
\end{eqnarray}
Using the population distribution $\overline{n_{I}(t)-n_{f}(t)}$
obtained for $T_{r}=T_{p}$, Eqs. (\ref{eq:25}) and (\ref{eq:26})
yield $\sigma_{1xx}\left(\omega\right)=\sigma_{oxx}\Theta(\omega-2\left|E_{F}\right|/\hbar)$
and $\sigma_{1yy}\left(\omega\right)=\sigma_{oyy}\Theta(\omega-2\left|E_{F}\right|/\hbar)$
with $\sigma_{oxx}=2.707\: e^{2}/\hbar$ and $\sigma_{oyy}=0.725\: e^{2}/\hbar$.
Using the population distribution $\overline{n_{I}(t)-n_{f}(t)}$
obtained for $T_{r}=T_{p}/10$, Eqs. (\ref{eq:25}) and (\ref{eq:26})
yield $\sigma_{1xx}\left(\omega\right)=\sigma_{oxx}\Theta(\omega-2\left|E_{F}\right|/\hbar)$
and $\sigma_{1yy}\left(\omega\right)=\sigma_{oyy}\Theta(\omega-2\left|E_{F}\right|/\hbar)$
with $\sigma_{oxx}=0.926\: e^{2}/\hbar$and $\sigma_{oyy}=-2.952\: e^{2}/\hbar$.
These results can be compared with the conductivity tensors obtained
in case of a graphene sheet in Ref.~\onlinecite{Ferreira}. The difference
here is that we have use the population distribution obtained by solving
the optical Bloch equations, whereas in Ref.~\onlinecite{Ferreira} the
Fermi Dirac distribution function has been used. 

Using Kramers-Kronig relation, $\sigma_{2}\left(\omega\right)$ can
be calculated from $\sigma_{1}\left(\omega\right)$ according to
\begin{equation}
\sigma_{2}\left(\omega\right)=-\frac{2}{\pi}\mathcal{P}\intop_{0}^{\infty}\frac{\omega\sigma_{1}\left(\omega^{'}\right)}{\omega^{'2}-\omega^{2}}\: d\omega^{'},\label{eq:28}
\end{equation}
where $\mathcal{P}$ denotes the Cauchy principle part of the integral. The
measurement of the Faraday rotation angle is performed with the probe
pulse with frequency in the transparency region. In the experiment,
the probe pulse has an energy of $2\hbar\omega_{F}-\hbar\delta$,
where $\hbar\omega_{F}$ is the Fermi energy and $\hbar\delta$ is
the detuning energy. Therefore the width of the transparency region
is given by $2\hbar\omega_{F}$. Thus, the Eq.~(\ref{eq:28}) can be
evaluated for $\hbar\omega\leq2\hbar\omega_{F}$. There are poles
at $\omega^{'}=\pm\omega$. Using $\sigma_{1xx}\left(\omega\right)$,
Eq.~(\ref{eq:28}) gives 
\begin{eqnarray}
\sigma_{2xx}\left(\omega\right) & = & -\frac{2\sigma_{oxx}}{\pi}\lim_{\eta\rightarrow0}\left[\intop_{0}^{\omega-\eta}\frac{\omega\Theta(\omega^{'}-2\omega_{F})}{\omega^{'2}-\omega^{2}}\: d\omega^{'} \right.
\nonumber\\
& & +\left.\intop_{\omega+\eta}^{\infty}\frac{\omega\Theta(\omega^{'}-2\omega_{F})}{\omega^{'2}-\omega^{2}}\: d\omega^{'}\right],\label{eq:29}
\end{eqnarray}
where $\eta$ is an infinitesimal positive quantity. Since $\hbar\omega<2\hbar\omega_{F}$,
the first integral in Eq.~(\ref{eq:29}) is zero. After evaluating the
second integral we get
\begin{equation}
\sigma_{2xx}\left(\omega\right)=i\sigma_{oxx}+\frac{2\sigma_{oxx}}{\pi}\arctan\mathrm{h}\left(\frac{2\omega_{F}}{\omega}\right)\label{eq:30}
\end{equation}
We are in the transparency region for the probe pulse, which means
$\frac{2\omega_{F}}{\omega}>1$. The function$\arctan\mathrm{h\left(z\right)}$
can be then expanded in terms of a Maclaurin series at infinity, i.e.
$\arctan\mathrm{h\left(z\right)}=-\frac{i\pi}{2}+\underset{n=1}{\overset{\infty}{\sum}}\frac{z^{-2n+1}}{2n-1}$.
Consequently, Eq.~(\ref{eq:30}) yields
\begin{equation}
\sigma_{2xx}\left(\omega\right)=\frac{2\sigma_{oxx}}{\pi}\underset{n=1}{\overset{\infty}{\sum}}\frac{1}{2n-1}\left(\frac{2\omega_{F}}{\omega}\right)^{-2n+1}.\label{eq:31}
\end{equation}
Similarly we obtain
\begin{equation}
\sigma_{2yy}\left(\omega\right)=\frac{2\sigma_{oyy}}{\pi}\underset{n=1}{\overset{\infty}{\sum}}\frac{1}{2n-1}\left(\frac{2\omega_{F}}{\omega}\right)^{-2n+1}.\label{eq:32}
\end{equation}

As shown in Ref.~\onlinecite{key-23-1}, there are interface bound states
(IBS) localized at two decoupled interface states of a PbTe/Pb$_{1-x}$Sn$_{x}$Te/PbTe
heterostructure with $d=10$ nm grown in the {[}111{]} direction.
It has been shown that the L-valley in {[}111{]} direction remains
gapless while gaps are opened in the oblique L valleys due to the
coupling of the IBS from the opposite interface states. Here we calculate
the Faraday rotation angle produced by the Weyl fermions at the two
interfaces with gapless L valley. We consider a structure with a slab
of thickness $d$ of 3D TI material Pb$_{1-x}$Sn$_{x}$Te sandwiched
by PbTe with thickness $t$, as shown in Fig. \ref{3DTI_slab}a. We
choose the thickness of the slab to be $d=10$ nm. A probe pulse linearly
polarized along the $x+y$-direction and propagating along $z$-direction
travels perpendicularly to the two interfaces. This probe pulse is
partially reflected and partially transmitted at the boundaries. Solutions
inside and outside the material can be solved by dividing the space
into five different regions as shown in Fig. \ref{3DTI_slab}b, where
$E_{I}$, $E_{II}$, $E_{III}$, $E_{IV}$ and $E_{V}$,are the fields
in the region $I$, $II$, $III$, $IV$ and $IV$, respectively.
The solutions are
\begin{eqnarray}
\mathbf{E}_{I} & = & \left[\begin{array}{c}
E_{ax}\\
E_{ay}
\end{array}\right]e^{ik_{I}z}+\left[\begin{array}{c}
E_{bx}\\
E_{by}
\end{array}\right]e^{-ik_{I}z}, 
\\
\mathbf{E}_{II} & = & \left[\begin{array}{c}
E_{cx}\\
E_{cy}
\end{array}\right]e^{ik_{II}z}+\left[\begin{array}{c}
E_{dx}\\
E_{dy}
\end{array}\right]e^{-ik_{II}z},
\\
\mathbf{E}_{III} & = & \left[\begin{array}{c}
E_{ex}\\
E_{ey}
\end{array}\right]e^{ik_{III}z}+\left[\begin{array}{c}
E_{fx}\\
E_{fy}
\end{array}\right]e^{-ik_{III}z},
\\
\mathbf{E}_{IV} & = & \left[\begin{array}{c}
E_{gx}\\
E_{gy}
\end{array}\right]e^{ik_{IV}z}+\left[\begin{array}{c}
E_{hx}\\
E_{hy}
\end{array}\right]e^{-ik_{II}z},
\\
\mathbf{E}_{V} & = & \left[\begin{array}{c}
E_{ix}\\
E_{iy}
\end{array}\right]e^{ik_{I}z},
\end{eqnarray}
where $E_{\alpha x}$ ($E_{\alpha y}$), $\alpha=a,b,c,d,e,f,g,h,i$,
are the $x$ ($y$) components of the field amplitudes in regions $I$ through $V$. $k_{I}$, $k_{II}$ and $k_{III}$ are the wave
vectors in air (region $I$), in PbTe (region $II$) and in Pb$_{1-x}$Sn$_{x}$Te
(region $III$), respectively. The incident probe pulse is polarized
along the $x+y$-axis. Therefore $E_{ax}=E_{ay}$. For simplicity,
we assume that the wave vectors within the material Pb$_{1-x}$Sn$_{x}$Te
and PbTe do not differ significantly and thus $k_{II}\thickapprox k_{III}$. 

Our geometry has a dimension of length $2t+d$ with top, bottom, and
interface surfaces being parallel to the plane of polarization. Rotation
of the polarization on the Poincare sphere is due to the charge carriers
at the interfaces, which are excited by the pump pulse with energy
at least twice the Dirac point energy measured from the Fermi level
(see Fig. \ref{OpticalExcitation}). The accumulation of the phase
difference is only due to surface carriers that come from the difference
in the optical conductivity tensor for the $x$ and $y$ polarization
of the light. There is no contribution to the phase shift in the polarization
from the bulk. However, the index of refraction of the bulk leads
to interference effects due to reflection and transmission at the
boundaries. The Maxwell equations to be solved are given by\cite{Ferreira}
\begin{eqnarray}
\frac{\partial^{2}E_{i}}{\partial z^{2}} & = & i\omega\mu_{o}\left[\delta\left(z-t\right)+\delta\left(z-t-d\right)\right]\underset{j=x,\, y}{\sum}\sigma_{ij}E_{j} \nonumber\\
& & +\omega^{2}\epsilon_{r}\mu_{o}E_{i},
\end{eqnarray}
where $\mu_{o}$ is the permeability of the free space and $\epsilon_{r}$
is the dielectric constant of the material in the bulk. It is important
to note that the delta functions ensure that the optical conductivity
tensor originates only from the interface carriers. The optical conductivity
tensor that enters Maxwell's equations is the imaginary part
of $\sigma\left(\omega\right)$, i.e. $\sigma_{2}\left(\omega\right)$
[see Eqs. (\ref{eq:31}) and (\ref{eq:32})], which gives rise to the dispersion
of the incident light inside the material. The boundary conditions
are determined by the continuity of the tangential components of the
electric field and their derivatives at the boundaries of the materials
at $z=0$, $z=t$, $z=t+d$ and $z=2t+d$. The details of matching
of the fields at the boundaries are shown in Appendix A. The transmission
amplitudes for $x$ and $y$ components of the electric field are
calculated to be $T_{x,y}=E_{ix,iy}/E_{a}=\left|T_{x,y}\right|e^{i\theta_{x,y}}$,
where $\left|T_{x,y}\right|$ is the transmission amplitude and $\theta_{x,y}$
are the Faraday rotation angles for the light polarized in $x$ and
$y$ direction. $T_{x}$ and $T_{y}$ are given by
\begin{eqnarray}
T_{x} & = & 4k_{I}k_{II}e^{-ik_{I}\left(2t+d\right)} \nonumber\\
& & /\left\{(k_{I}+k_{II})e^{-ik_{II}t}\left[\alpha(k_{II}A_{x}+C_{x})\right.\right. \nonumber\\
& & +\left.\beta(k_{II}B_{x}+D_{x})\right]+(k_{I}-k_{II})e^{ik_{II}t} \nonumber\\
& & \times\left.\left[\alpha(k_{II}A_{x}-C_{x})+\beta(k_{II}B_{x}-D_{x})\right]
\right\}\label{eq:33}
\\
T_{y} & = & 4k_{I}k_{II}e^{-ik_{I}\left(2t+d\right)} \nonumber\\
& & /\left\{(k_{I}+k_{II})e^{-ik_{II}t}\left[\alpha(k_{II}A_{y}+C_{y})\right.\right. \nonumber\\
& & +\left.\beta(k_{II}B_{y}+D_{y})\right]+(k_{I}-k_{II})e^{ik_{II}t} \nonumber\\
& & \times\left.\left[\alpha(k_{II}A_{y}-C_{y})+\beta(k_{II}B_{y}-D_{y})\right]\right\}\label{eq:34}
\end{eqnarray}
where $A_{x}$($A_{y}$) , $B_{x}$($B_{y}$), $C_{x}$($C_{y}$)
and $D_{x}$($D_{y}$) are the $x$($y$) components of the parameters
$A$, $B$, $C$ and $D$, respectively (see Appendix A). After solving
Eqs. \ref{eq:33} and \ref{eq:34} for $\theta_{x}$ and $\theta_{y}$,
we write the Faraday rotation angle as $\theta_{F}=\left(\theta_{x}-\theta_{y}\right)/2$.
The useful quantity, the total transmittance$\mathrm{\top}$, which
measures the energy of the electromagnetic field inside the material,
can be defined as $\top=\left(\left|T_{x}\right|^{2}+\left|T_{y}\right|^{2}\right)/2$.

\begin{figure}

\includegraphics[width=8.5cm]{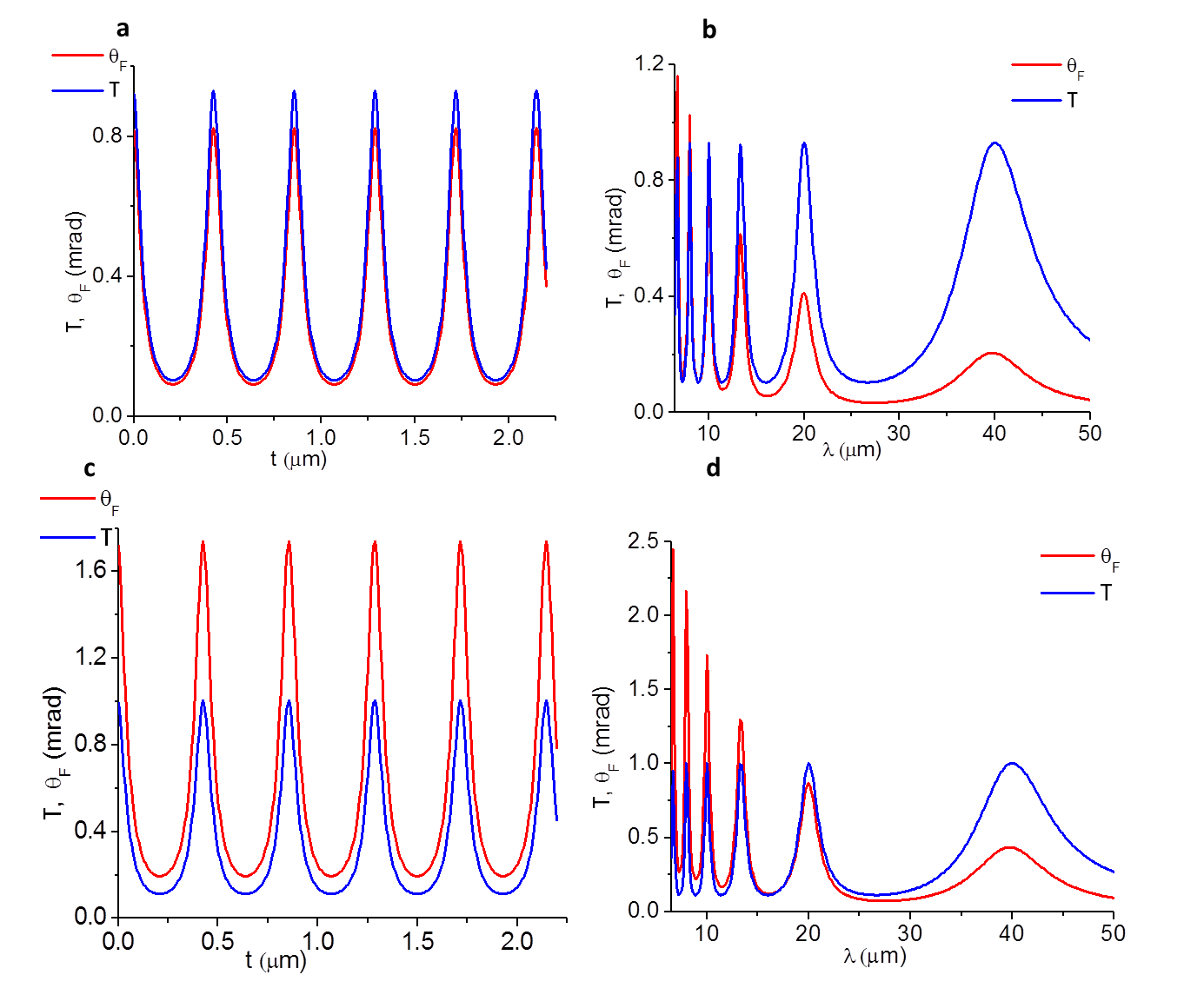}\caption{Transmittance and the Faraday rotation angle are plotted as a function
of thickness $t$ and as a function of wavelength $\lambda$ for the
geometry shown in Fig. \ref{3DTI_slab}. In \textbf{a} and \textbf{b}
we choose $T_{r}=T_{p}=1$ ps, while in \textbf{c} and \textbf{d}
we choose $T_{r}=T_{p}/10=1$ ps. For \textbf{a} and \textbf{c} the
wavelength is $\lambda=9.97\:\mu m$, which corresponds to a detuning
energy of 10 meV. For \textbf{b} and \textbf{d} the thickness is $t=1.72\:\mu m$.
The width of the transparency region of the excitation of Weyl fermion
is 134 meV, as calculated from the bandstructure as shown in Fig.
\ref{BandStructure}.
\label{FaradayAngle&Transmittance}}

\end{figure}

From the bandstructure calculation we obtain that the Fermi level
lies around 67 meV below the Dirac point. Therefore, we choose a transparency
energy gap of $\hbar\omega_{cv}=2\left|E_{F}\right|$, which is 134
meV in our calculation. A linearly polarized probe pulse with detuning
energy of $\hbar\delta=10$ meV, pulse duration of 1 ps and bandwidth
of $\hbar\gamma=4$ meV can be used. In Figs.~\ref{FaradayAngle&Transmittance}
\textbf{a} and \textbf{b} we show the transmittance and the Faraday
rotation angle for $T_{r}=T_{p}$. In Figs. \ref{FaradayAngle&Transmittance}
\textbf{c} and \textbf{d} the transmittance and the Faraday rotation
for $T_{r}=T_{p}/10$ are shown. For the transmittance and the Faraday
rotation angle as a function of thickness $t$ the wavelength is chosen
to be $\lambda=9.97\:\mu m$. For the transmittance and the Faraday
rotation angle as a function of wavelength $\lambda$ the thickness
of PbTe layers is taken to be $t=1.720\;\mu m$. It is seen from the
figures that the Faraday rotation angle follows exactly the transmittance.
In particular, the maxima of the Faraday rotation angle occur at the
maxima of the transmittance, which corresponds the case of nearly
reflectionless slab in optics. There are two cases when reflection
turns to zero. The first case is given by the half-wave condition
when $w=m\lambda/2n$, $n$ is an integer and $n_{1}=n_{2}$. The
second case is given by the quarter-wave condition when $w=\left(2m+1\right)\lambda/4n$,
$n=\sqrt{n_{1}n_{2}}$, where $w$ is the total length of the slab, $m$ is an integer, $n$, $n_{1}$ and $n_{2}$ are the indices
of refraction of a slab of material and of the materials on either
side of the slab, respectively. In our case the half-wave condition
is met. Therefore, the resonances are seen (Fig.~\ref{FaradayAngle&Transmittance}\textbf{b})
inside the material at half-integer multiples of the probe wavelength
divided by the index of refraction of the material, which is $n=5.8$.
Of course, Fig.~\ref{FaradayAngle&Transmittance} exhibits a slight
deviation from zero reflection at maxima due to the presence of multiple
interfaces. The Faraday rotation angle obtained using a wide-bandgap
semiconductor quantum dot is usually small compared to this result.
\cite{Berezovsky}

\section{Conclusion}

We have calculated the optical transitions for the Weyl interface
fermions in 3D TI at the $L$ point using the Dirac Hamiltonian. The
spin selection rules for the optical transitions are very strict.
The interaction Hamiltonian that comes from the quadratic part of
the \textbf{k}.\textbf{ p} is included in the calculation and is shown
to have zero contribution to the transition dipole moment. 

We demonstrate the effect of the strict optical selection rules by
considering the Faraday effect due to Pauli exclusion principle in
a pump-probe setup. Our calculations show that the Faraday rotation
angle exhibits oscillations as a function of probe wavelength and
thickness of the slab material on either side of the 3D TI double
interface of a PbTe/Pb$_{0.31}$Sn$_{0.69}$Te/PbTe heterostructure.
The maxima in the Faraday rotation angle are in the millirad regime. 
\begin{acknowledgments}
We acknowledge support from NSF (Grant ECCS-0901784), AFOSR (Grant
FA9550-09-1-0450), and NSF (Grant ECCS-1128597). We thank Gerson Ferreira
for useful discussions. M.N.L. thanks Daniel Loss for fruitful discussions
during his stay at the University of Basel, Switzerland. M.N.L. acknowledges
partial support from the Swiss National Science Foundation.
\end{acknowledgments}

\section*{Appendix A}

The continuity of the tangential components of the electric field
at $z=0$, $z=t$, $z=t+d$ and $z=2t+d$ leads to
\begin{eqnarray}
\lefteqn{
\left[\begin{array}{c}
E_{a}\\
E_{a}
\end{array}\right]+\left[\begin{array}{c}
E_{bx}\\
E_{by}
\end{array}\right]=\left[\begin{array}{c}
E_{cx}\\
E_{cy}
\end{array}\right]+\left[\begin{array}{c}
E_{dx}\\
E_{dy}
\end{array}\right],
}
\label{eq:36}
\\
\lefteqn{
\left[\begin{array}{c}
E_{cx}\\
E_{cy}
\end{array}\right]e^{ik_{II}t}+\left[\begin{array}{c}
E_{dx}\\
E_{dy}
\end{array}\right]e^{-ik_{II}t}=\left[\begin{array}{c}
E_{ex}\\
E_{ey}
\end{array}\right]e^{ik_{II}t}
}
\nonumber\\ 
& & +\left[\begin{array}{c}
E_{ex}\\
E_{ey}
\end{array}\right]e^{-ik_{II}t},
\\
\lefteqn{
\left[\begin{array}{c}
E_{ex}\\
E_{ey}
\end{array}\right]e^{ik_{II}\left(t+d\right)}+\left[\begin{array}{c}
E_{fx}\\
E_{fy}
\end{array}\right]e^{-ik_{II}\left(t+d\right)}
}
\nonumber\\
& & =\left[\begin{array}{c}
E_{gx}\\
E_{gy}
\end{array}\right]e^{ik_{II}\left(t+d\right)}+\left[\begin{array}{c}
E_{hx}\\
E_{hy}
\end{array}\right]e^{-ik_{II}\left(t+d\right)},
\\
\lefteqn{ 
\left[\begin{array}{c}
E_{gx}\\
E_{gy}
\end{array}\right]e^{ik_{II}\left(2t+d\right)}+\left[\begin{array}{c}
E_{hx}\\
E_{hy}
\end{array}\right]e^{-ik_{II}\left(2t+d\right)}
}
\nonumber\\
& & =\left[\begin{array}{c}
E_{ix}\\
E_{iy}
\end{array}\right]e^{ik_{I}\left(2t+d\right)}.
\end{eqnarray}
Similarly the continuity of derivative of the electric fields at $z=0$,
$z=t$, $z=t+d$ and $z=2t+d$ yields
\begin{widetext}
\begin{eqnarray}
\lefteqn{
ik_{I}\left[\begin{array}{c}
E_{a}\\
E_{a}
\end{array}\right]-ik_{I}\left[\begin{array}{c}
E_{bx}\\
E_{by}
\end{array}\right]=ik_{II}\left[\begin{array}{c}
E_{cx}\\
E_{cy}
\end{array}\right]-ik_{II}\left[\begin{array}{c}
E_{dx}\\
E_{dy}
\end{array}\right],
}
\label{eq:40}
\\
\lefteqn{
ik_{II}\left[\begin{array}{c}
E_{cx}\\
E_{cy}
\end{array}\right]e^{ik_{II}t}-ik_{II}\left[\begin{array}{c}
E_{dx}\\
E_{dy}
\end{array}\right]e^{-ik_{II}t}
=ik_{II}\left[\begin{array}{c}
E_{ex}\\
E_{ey}
\end{array}\right]e^{ik_{II}t}-ik_{II}\left[\begin{array}{c}
E_{fx}\\
E_{fy}
\end{array}\right]e^{-ik_{II}t}
}
\nonumber\\
& & +i\omega\mu_{o}\left[\begin{array}{c}
\sigma_{xx}(E_{ex}e^{ik_{II}t}-E_{fx}e^{-ik_{II}t})+\sigma_{xy}(E_{ey}e^{ik_{II}t}-E_{fy}e^{-ik_{II}t})\\
\sigma_{yx}(E_{ex}e^{ik_{II}t}-E_{fx}e^{-ik_{II}t})+\sigma_{yy}(E_{ey}e^{ik_{II}t}-E_{fy}e^{-ik_{II}t})
\end{array}\right],
\label{eq:41}
\\
\lefteqn{
ik_{II}\left[\begin{array}{c}
E_{ex}\\
E_{ey}
\end{array}\right]e^{ik_{II}\left(t+d\right)}-ik_{II}\left[\begin{array}{c}
E_{fx}\\
E_{fy}
\end{array}\right]e^{-ik_{II}\left(t+d\right)}
=ik_{II}\left[\begin{array}{c}
E_{gx}\\
E_{gy}
\end{array}\right]e^{ik_{II}\left(t+d\right)}-ik_{II}\left[\begin{array}{c}
E_{hx}\\
E_{hy}
\end{array}\right]e^{-ik_{II}\left(t+d\right)}
}
\nonumber\\
& & +i\omega\mu_{o}\left[\begin{array}{c}
\sigma_{xx}(E_{gx}e^{ik_{II}\left(t+d\right)}-E_{hx}e^{-ik_{II}\left(t+d\right)})+\sigma_{xy}(E_{gy}e^{ik_{II}\left(t+d\right)}-E_{hy}e^{-ik_{II}\left(t+d\right)})\\
\sigma_{yx}(E_{gx}e^{ik_{II}\left(t+d\right)}-E_{hx}e^{-ik_{II}\left(t+d\right)})+\sigma_{yy}(E_{gy}e^{ik_{II}\left(t+d\right)}-E_{gy}e^{-ik_{II}\left(t+d\right)})
\end{array}\right],
\label{eq:42}
\\
\lefteqn{
ik_{II}\left[\begin{array}{c}
E_{gx}\\
E_{gy}
\end{array}\right]e^{ik_{II}\left(2t+d\right)}-ik_{II}\left[\begin{array}{c}
E_{hx}\\
E_{hy}
\end{array}\right]e^{-ik_{II}\left(2t+d\right)}
=ik_{I}\left[\begin{array}{c}
E_{ix}\\
E_{iy}
\end{array}\right]e^{ik_{I}\left(2t+d\right)}.
}
\label{eq:43}
\end{eqnarray}
\end{widetext}
The response of the top and bottom surfaces of the Pb$_{1-x}$Sn$_{x}$Te
slab to the field depends on the transition matrix elements on the
corresponding surfaces. Since, the transition matrix elemets for both
of the surfaces are same, we have $\sigma_{t,ij}=\sigma_{t+L,ij}$.
The off diagonal elements $\sigma_{xy}$ and $\sigma_{xy}$ of the
magneto-optical tensors $\sigma_{ij}$ are calculated to be zero.
The algebric Eqs. \ref{eq:36} to \ref{eq:43} can be solved for each
of the amplitude of component field in each region interm of the incident
filed. The solutions for the transmitted field are given by 
\begin{widetext}
\begin{equation}
\left[\begin{array}{c}
E_{ix}\\
E_{iy}
\end{array}\right]=\frac{4k_{I}k_{II}e^{-ik_{I}\left(2t+d\right)}}{(k_{I}+k_{II})e^{-ik_{II}t}\left[\alpha(k_{II}A+C)+\beta(k_{II}B+D)\right]+(k_{I}-k_{II})e^{ik_{II}t}\left[\alpha(k_{II}A-C)+\beta(k_{II}B-D)\right]}\left[\begin{array}{c}
E_{a}\\
E_{a}
\end{array}\right],
\end{equation}
where
$\alpha=\frac{k_{II}+k_{I}}{4k_{II}}$,
$\beta=\frac{k_{II}-k_{I}}{4k_{II}}$,
\begin{eqnarray}
A & = & 2e^{-ik_{II}(t+d)}-\frac{2i\omega\mu_{o}\sin k_{II}d}{k_{II}}\left[\begin{array}{c}
\sigma_{xx}\\
\sigma_{yy}
\end{array}\right]e^{-ik_{II}t},
\\
B & = & 2e^{ik_{II}(t+d)}+\frac{2i\omega\mu_{o}\sin k_{II}d}{k_{II}}\left[\begin{array}{c}
\sigma_{xx}\\
\sigma_{yy}
\end{array}\right]e^{ik_{II}t},
\\
C & = & e^{-ik_{II}(t+d)}\left(k_{II}+\omega\mu_{o}\left[\begin{array}{c}
\sigma_{xx}\\
\sigma_{yy}
\end{array}\right]\right)\left(2+\frac{\omega\mu_{o}}{k_{II}}\left[\begin{array}{c}
\sigma_{xx}\\
\sigma_{yy}
\end{array}\right]\right)
 -\frac{\omega\mu_{o}}{k_{II}}e^{-ik_{II}(t+d)}\left[\begin{array}{c}
\sigma_{xx}\\
\sigma_{yy}
\end{array}\right]\left(-k_{II}-\omega\mu_{o}\left[\begin{array}{c}
\sigma_{xx}\\
\sigma_{yy}
\end{array}\right]\right),
\\
D & = & -\frac{\omega\mu_{o}}{k_{II}}e^{ik_{II}(t-d)}\left[\begin{array}{c}
\sigma_{xx}\\
\sigma_{yy}
\end{array}\right]\left(k_{II}+\omega\mu_{o}\left[\begin{array}{c}
\sigma_{xx}\\
\sigma_{yy}
\end{array}\right]\right)
-e^{-ik_{II}(t+d)}\left(-k_{II}-\omega\mu_{o}\left[\begin{array}{c}
\sigma_{xx}\\
\sigma_{yy}
\end{array}\right]\right)
\left(2+\frac{\omega\mu_{o}}{k_{II}}\left[\begin{array}{c}
\sigma_{xx}\\
\sigma_{yy}
\end{array}\right]\right).
\end{eqnarray}
\end{widetext}

\end{document}